\documentclass[aps,twocolumn,prl,preprintnumbers,amsmath,amssymb,superscriptaddress,floatfix]{revtex4}

\usepackage{graphicx}
\usepackage{bm}
\usepackage{hyperref}
\usepackage{natbib}
\usepackage{braket}
\usepackage{float}
\usepackage{amsmath}
\usepackage{color}

\begin{document}

\title{Asymmetric splitting of an antiferromagnetic resonance via quartic exchange interactions in multiferroic hexagonal HoMnO$_3$}

\author{N. J. Laurita}
\affiliation{The Institute for Quantum Matter, Department of Physics and Astronomy, The Johns Hopkins University, Baltimore, MD 21218, USA}

\author{Yi Luo}
\affiliation{The Institute for Quantum Matter, Department of Physics and Astronomy, The Johns Hopkins University, Baltimore, MD 21218, USA}

\author{Rongwei Hu}
\affiliation{Rutgers Center For Emergent Materials, Department of Physics and Astronomy, Rutgers University, Piscataway, NJ, 08854, USA}

\author{Meixia Wu}
\affiliation{Rutgers Center For Emergent Materials, Department of Physics and Astronomy, Rutgers University, Piscataway, NJ, 08854, USA}

\author{S. W. Cheong}
\affiliation{Rutgers Center For Emergent Materials, Department of Physics and Astronomy, Rutgers University, Piscataway, NJ, 08854, USA}

\author{O. Tchernyshyov}
\affiliation{The Institute for Quantum Matter, Department of Physics and Astronomy, The Johns Hopkins University, Baltimore, MD 21218, USA}

\author{N. P. Armitage}
\affiliation{The Institute for Quantum Matter, Department of Physics and Astronomy, The Johns Hopkins University, Baltimore, MD 21218, USA}

\date{\today}

\begin{abstract}
\noindent The \textit{symmetric} splitting of two spin-wave branches in an antiferromagnetic resonance (AFR) experiment has been an essential measurement of antiferromagnets for over half a century.  In this work, circularly polarized time-domain THz spectroscopy experiments performed on the low symmetry multiferroic h-HoMnO$_3$ reveal an AFR of the Mn sublattice to split \textit{asymmetrically} in applied magnetic field, with an $\approx$ 50\% difference in $g$-factors between the high and low energy branches of this excitation.  The temperature dependence of the $g$-factors, including a drastic renormalization at the Ho spin ordering temperature, reveals this asymmetry to unambiguously stem from Ho-Mn interactions.  Theoretical calculations demonstrate the AFR asymmetry is not explained by conventional Ho-Mn exchange mechanisms alone and are only reproduced if quartic spin interactions are also included in the spin Hamiltonian.  Our results provide a paradigm for the optical study of such novel interactions in hexagonal manganites and low symmetry antiferromagnets in general.
\end{abstract}

\maketitle

Antiferromagnetic resonance (AFR) has been perhaps the most essential property of antiferromagnets since the earliest description by Kittel over half a century ago \cite{Kittel1951}.  In an AFR experiment, two spin-wave branches, each active to a different helicity of circularly polarized light, symmetrically split in applied magnetic field.  However, changes to this phenomena may occur in low symmetry environments, as interactions between localized spins in magnetic insulators are heavily influenced by the symmetry of the crystal structure in which they are embedded.  The hexagonal rare-earth manganites h-RMnO$_3$ are prime examples of materials whose low symmetry results in remarkable physical behavior \cite{Lorenz2013}, including multiferroism and exceptionally strong magnetoelectric coupling \cite{Hur2009}.  Magnetism in these systems consists of both rare-earth and manganese magnetic moments, which lie in orthogonal directions due to crystalline anisotropy \cite{Nandi2008}.  Interactions between these moments has been a topic of intense investigation \cite{Fiebig2001, Fabreges2008, Talbayev2008, Nandi2008, Meier2012} as such couplings are thought to drive magnetic transitions \cite{Condran2010} and mediate magnetoelectric phenomena \cite{Lottermoser2004, Ueland2010}.  However, the exchange mechanism between R-Mn spins has remained elusive, as their orthogonality suggests a less conventional interaction than Heisenberg exchange in the spin equilibrium configuration.

\begin{figure}
\includegraphics[width=0.82\columnwidth, keepaspectratio]{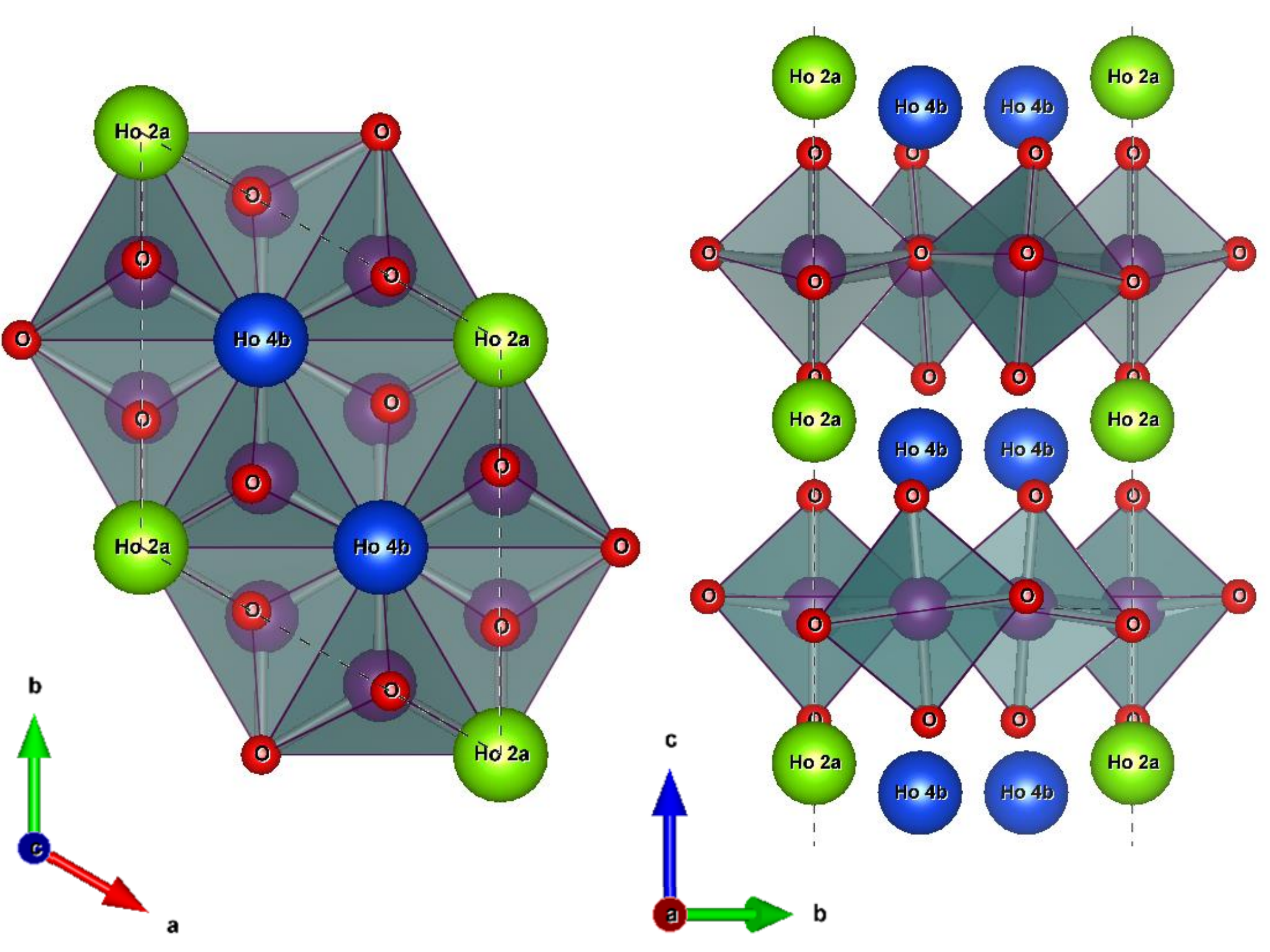}
\caption{Crystal structure of HMO in the ferroelectric phase (T $<$ T$_{c}$=875K) with views along the (left) $c$ axis and (right) $a$ axis respectively.  In this phase, the Ho$^{+3}$ ions (green and blue spheres) lie in symmetry distinct positions of the lattice resulting in a finite ferroelectric moment along the $c$ axis.}
\label{Fig0}
\end{figure} 

Of these materials, h-HoMnO$_3$ (HMO) possesses the largest effective rare-earth magnetic moment and is thus ideal for studying magnetic exchange in these systems \cite{Hur2009, Lorenz2004A, Cruz2005, Nobuyuki1998, Yen2005}.  The hexagonal crystal structure of HMO (Figure \ref{Fig0}) consists of alternating layers of corner sharing MnO$_5$ bipyramids and Ho ions which are stacked along the $c$ axis \cite{Momma2011}.  At the ferroelectric transition, T$_c$=875K, the MnO$_5$ bipyramids buckle \cite{Cheong2007, VanAken2004, Fennie2005} reducing the symmetry to the non-centrosymmetric polar space group P6$_3$cm with Ho ions occupying two symmetry distinct positions of the crystal lattice.  The S$_\text{Mn}$=2 spins form a two-dimensional frustrated triangular lattice which orders at T$_\text{N}$ $\approx$ 75K in a 120 degree structure with symmetry P6$^{'}_3$c$^{'}$m \cite{Fiebig2000, Fiebig2002, Fiebig2003, Vajk2005, Brown2006, Brown2008}.  Two additional zero field Mn sublattice transitions occur at T$_{\text{SR}}$ $\approx$ 40K (P6$^{'}_3$cm$^{'}$) and at T$_{\text{Ho}}$ $\approx$ 5K (P6$_3$cm), in which the Mn spins rotate by 90 degrees within the basal plane.  The ordering of the Ho sublattices is less understood  \cite{Nandi2008, Sugie2002, Fiebig2002, Fiebig2000, Fiebig2003, Vajk2005, Lonkai2002, Brown2006, Brown2008}, however it is expected that the S$_\text{Ho}$=2 spins order antiferromagnetically along the $c$ axis due to uniaxial anisotropy \cite{Nandi2008}, with experimental evidence suggesting that magnetization of at least one of the Ho sublattices onsets near T$_{\text{SR}}$ \cite{Lonkai2002, Nandi2008} and some form of long range order existing  below T$_\text{Ho}$.  

Interactions between R and Mn moments in hexagonal manganites can be probed by examining the spin excitations of the Mn sublattice, whose minimal spin Hamiltonian is given by:
\begin{equation}
H = J \sum _{<ij>} \mathbf S_i \cdot \mathbf S_j + \Delta \sum_{i} ({S_{i}^{z})}^2 - g \mu_B \mathbf B \cdot \sum _{i} \mathbf S_{i},
\label{MNSpinHam}
\end{equation}
where $J$ is the Heisenberg exchange, $\Delta$ is the planar anisotropy, $\mathbf B \parallel c$ is the applied magnetic field, $g$ = 2 is the Mn $g$-factor, and the sum is over neighboring pairs \cite{Palme1990, Vajk2005}.  The ground state of the Mn sublattice is a 120$^\circ$ ordered AF.  In the $\vec{k} \rightarrow 0$ limit (applicable to our optical measurements) the low energy spectrum consists of a Goldstone mode and a gapped AFR \cite{Palme1990}.  In the weak field limit, valid for fields $H$ $<$ $S_\text{Mn}J \approx 40$T in HMO, the energies of the AFR are given by:
\begin{equation}
\hbar \omega _{\pm}(B) = \hbar \omega(0) \pm g_\mathrm{eff} \mu _B B, 
\quad
\hbar \omega(0) = 3S\sqrt{J\Delta},
\label{LowHAFR}
\end{equation}
revealing two modes which split symmetrically in field with $g_\mathrm{eff} = \frac{g}{ 2+4\Delta/9J }$ (see Eq's 8-12 of Sec. III in the SI for derivation) \cite{Standard2012}.  Note that even for small anisotropy, $g_\mathrm{eff}$ is approximately half the bare ionic value. This is a particular feature of the 120$^\circ$ ordered AF which arises due to the low symmetry of the ordered state resulting in a not well defined $z$ angular momentum quantum number.  With exchange and anisotropy found to be $J$ = 2.44 meV and $\Delta$ = 0.38 meV respectively in HMO \cite{Vajk2005}, one expects $g_\mathrm{eff} = 0.97$ from \autoref{LowHAFR}.  However, in actuality much larger $g$-factors are observed at low temperatures in hexagonal manganites \cite{Standard2012, Talbayev2008}.  This has been explained by introducing an additional Heisenberg exchange interaction which ferromagnetically couples R spins to the finite $S_z$ of the Mn AFR modes into \autoref{MNSpinHam} \cite{Talbayev2008}.  However, such a coupling is expected to vanish in the ground state due to the orthogonality of spins, leaving the dominant equilibrium R-Mn spin interaction unresolved.

\begin{figure}
\includegraphics[width=1.0\columnwidth, keepaspectratio]{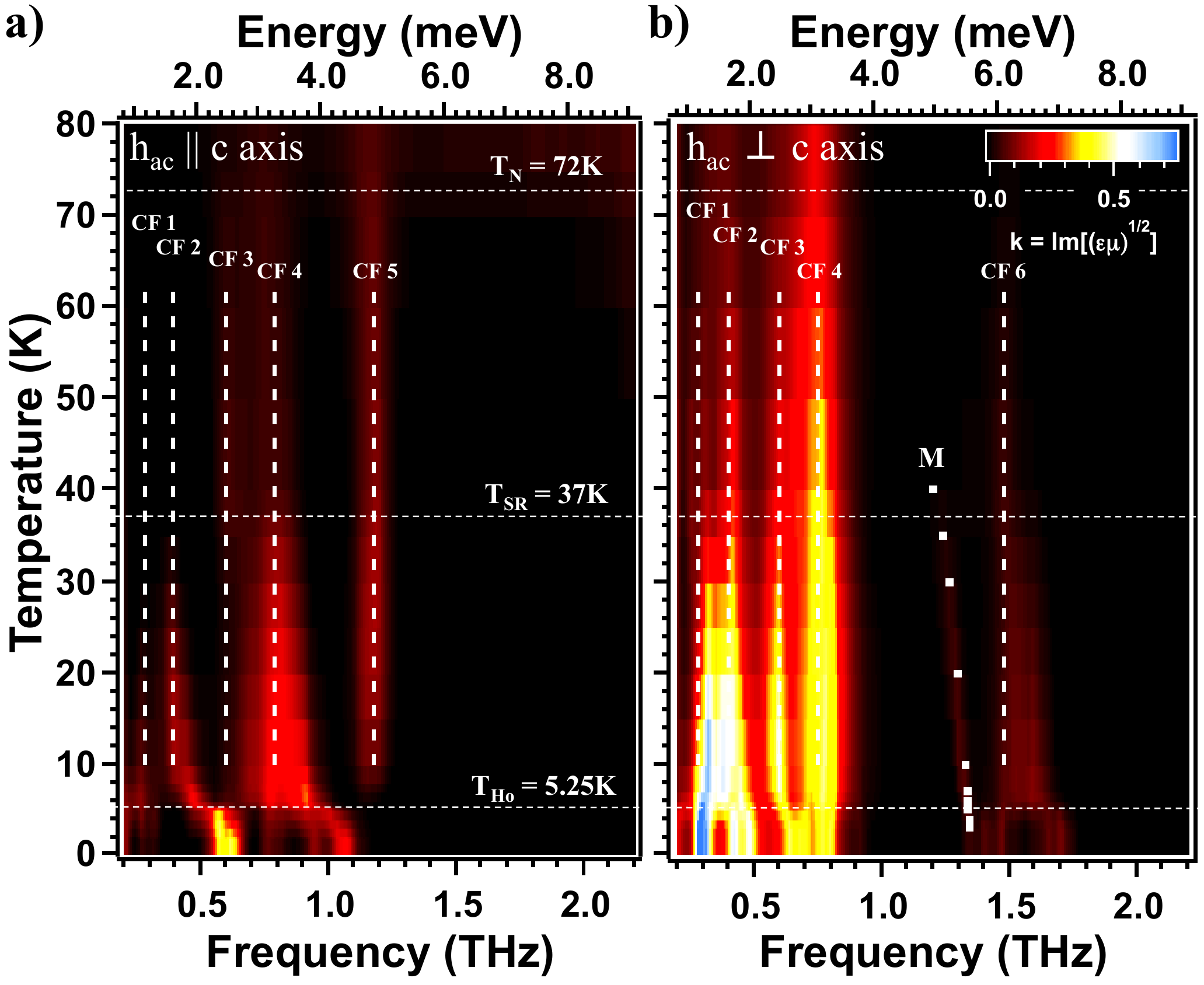}
\caption{Image plots of the imaginary part of the index of refraction as a function of temperature and frequency for the orientations (a) $\vec{\text{h}}_{\text{ac}}$ $\parallel$ c and (b) $\vec{\text{h}}_{\text{ac}}$ $\perp$ c respectively.  Horizontal dashed lines denote the three zero field transition temperatures while vertical dashed lines label the more prominent Ho crystal field excitations identified at temperatures T $\geq$ T$_{\text{Ho}}$.  The excitation labeled "M" is the AFR of the Mn sublattice whose resonant frequency is marked by white squares.}
\label{Fig1}
\end{figure} 

In this Letter, we present a systematic study of the low energy optical response of HMO via high resolution time-domain terahertz spectroscopy (TDTS).  We demonstrate that the Mn AFR possesses distinct selection rules to circularly polarized light, which allows our experiments to resolve the field dependent splitting of the AFR in weak magnetic fields with high precision.  The AFR is found to unexpectedly split \textit{asymmetrically} in magnetic field.  Careful study of the temperature dependence of this asymmetry unambiguously demonstrates the effect to stem from R-Mn interactions.  Theoretical investigations concludes the asymmetry is not explained by conventional R-Mn exchange mechanisms alone and is only reproduced if novel quartic spin interactions are also included in the spin Hamiltonian.  The potential for such interactions in other hexagonal manganites is discussed.

Single crystals of HMO were grown via optical floating zone method.  Two samples with the orientations [-1,1,0] (d = 670 $\mu$m) and [0,0,1] (d =  590 $\mu$m) normal to the sample surface were measured in this study.  TDTS transmission experiments were performed using a home-built spectrometer \cite{Laurita2016} in magnetic fields up to 6T in Faraday geometry ($\vec{k}_\text{THz}$ $\parallel$ $\vec{\text{H}}_{\text{dc}}$).  Via a coupling of the THz fields to both electric and magnetic dipole transitions of the sample, TDTS accesses the sample's electromagnetic response with exceptional resolution from 0.2 - 2 THz.  

\begin{figure*}
\includegraphics[width=1.75\columnwidth, keepaspectratio]{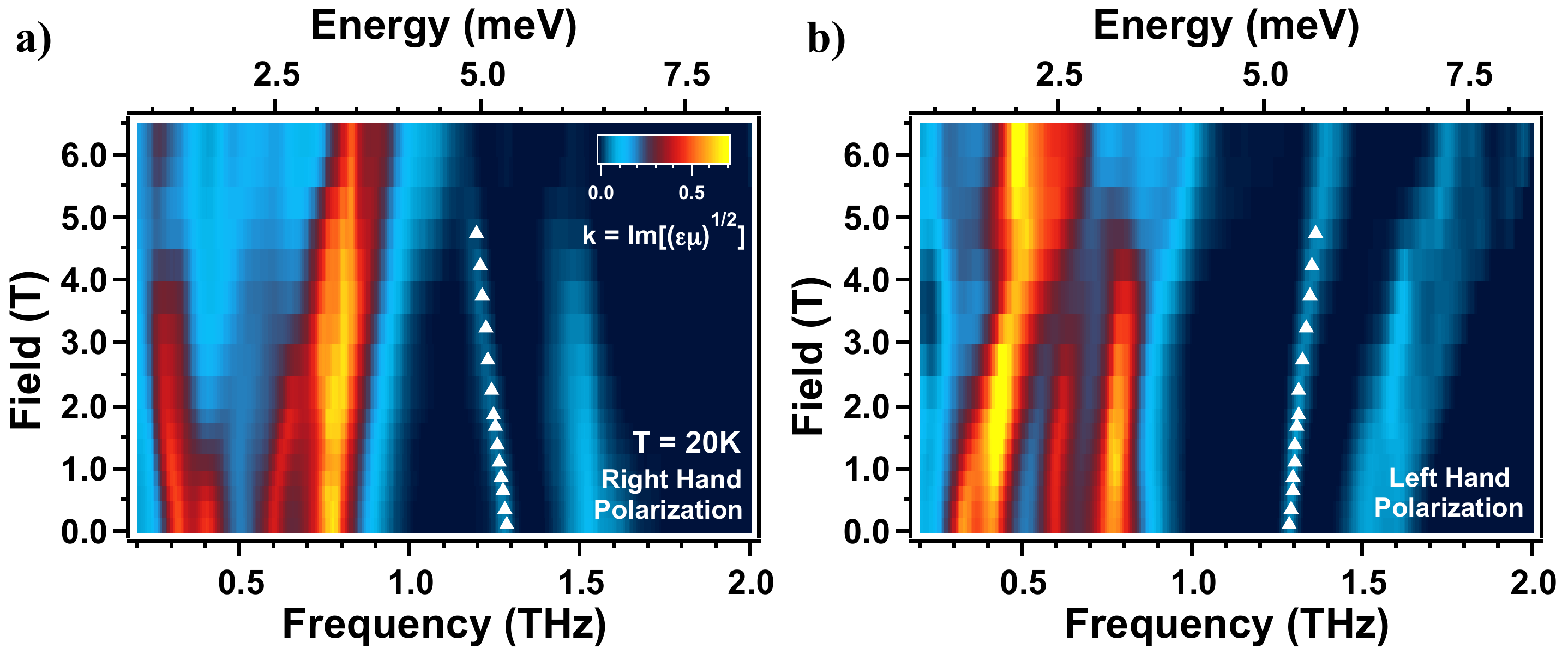}
\caption{Field dependence of the imaginary part of the index of refraction of HMO at 20K for (a) right and (b) left hand circular polarizations with $H$ $\parallel$ $c$.  The Mn AFR is the linearly varying excitation at $\approx$ 1.3 THz, which can be seen to naturally partition into low and high energy branches in the circular basis.  White triangles mark the extracted resonant frequencies of the AFR.}
\label{Fig2}
\end{figure*}

\autoref{Fig1} displays image plots of the imaginary, or dissipative, part of the complex index of refraction, $\tilde{n}$ = $n$ + $ik$, of HMO for the orientations (a) $\vec{\text{h}}_{\text{ac}}$ $\parallel$ c and (b) $\vec{\text{h}}_{\text{ac}}$ $\perp$ c respectively (full data set in Sec. IV of the  SI). One can show that the axial symmetry of the lattice constrains the zero field linear response such that only these two orientations give unique responses \cite{Armitage2014}.  The spectra are in excellent agreement with previous studies \cite{Talbayev2008, Bowlan2016}.  Many of the features seen in \autoref{Fig1} can be attributed to crystal field transitions of the Ho$^{+3}$ ($^5$I$_8$) ions, which have been previously discussed in the context of a number of compounds \cite{Judd1955, Abragam1970, Elliott1953, Rosenkranz2000, Ranon1969, Sirenko2008, Stevens1952}.  Several of the more prominent crystal field levels are labeled in \autoref{Fig1} and discussed in detail in Sec. IV of the SI (see Table 1).  Abrupt changes in the spectra, including a previously undiscovered dramatic renormalization of the crystal field excitation energies at $\approx$ 5K, identify the three zero field magnetic transitions at T$_{\text{N}}$ = 72K, T$_{\text{SR}}$ = 37K, and T$_{\text{Ho}}$ = 5.25K.  Here we focus on the AFR of the Mn sublattice which is labeled ``M" in \autoref{Fig1}.  In order to extract the dynamical properties of this mode, the spectra were fit with a Drude-Lorentz oscillator on a linear background to account for neighboring crystal field levels.  White squares in \autoref{Fig1} mark the extracted resonant frequencies of the Mn AFR. 

Measurements were then performed as a function of magnetic field to investigate the field dependent splitting of the AFR.  The hexagonal symmetry of HMO along with the $\mathcal{T}$ symmetry breaking under applied field constrains the linear response transmission matrix \cite{Jones1941, Armitage2014} such that it must be fully antisymmetric in the linear basis:
\begin{equation}
\tilde{T}_\text{linear}
=
\begin{bmatrix}
\tilde{T}_{xx} & \tilde{T}_{xy} \\ -\tilde{T}_{xy} & \tilde{T}_{xx}
\end{bmatrix}
\end{equation}
Such a fully antisymmetric transmission matrix can be diagonalized by a circular basis transformation as: 
\begin{equation}
\tilde{T}_\text{circular}
=
\begin{bmatrix}
\tilde{T}_{xx} + i \cdot \tilde{T}_{xy} & 0 \\ 0 & \tilde{T}_{xx} - i \cdot \tilde{T}_{xy}
\end{bmatrix}
=
\begin{bmatrix} \tilde{T}_{r}&0 \\ 0 & \tilde{T}_{l} \end{bmatrix}
\end{equation}
where $\tilde{T}_l$ and $\tilde{T}_r$ refer to the transmission of left and right hand circularly polarized light, the eigenpolarizations, respectively.  The above analysis suggests that experiments performed in Faraday geometry are best understood in the circular basis (see Sec. I of the SI for further details).  TDTS measurements performed here utilized a rotating polarizer technique, which allows for measurement of the sample's response to two polarization directions simultaneously and thus conversion to the circular basis \cite{Morris2012}.  

\autoref{Fig2}(a,b) displays image plots of the dissipative part of the index of refraction as a function of magnetic field at 20K for right and left hand circular polarizations respectively.  The excitation  at $\approx$ 1.3 THz which linearly varies with magnetic field is the AFR of the Mn sublattice.  One can immediately see that the two branches of the AFR possess distinct selection rules to right and left hand circular polarizations.  Such a partitioning of the AFR allows unique access to the splitting of this mode in weak magnetic fields, within the low field ``intermediate" phase of HMO, where the two branches would otherwise be highly overlapping in the linear basis.  In a similar manner as the zero field data, these spectra were fit to extract the magnetic field dependent dynamical properties of the AFR.  White triangles in \autoref{Fig2} mark the resonant frequency of the AFR at fields in which it is well defined.   

\begin{figure}
\includegraphics[width=0.92\columnwidth, keepaspectratio]{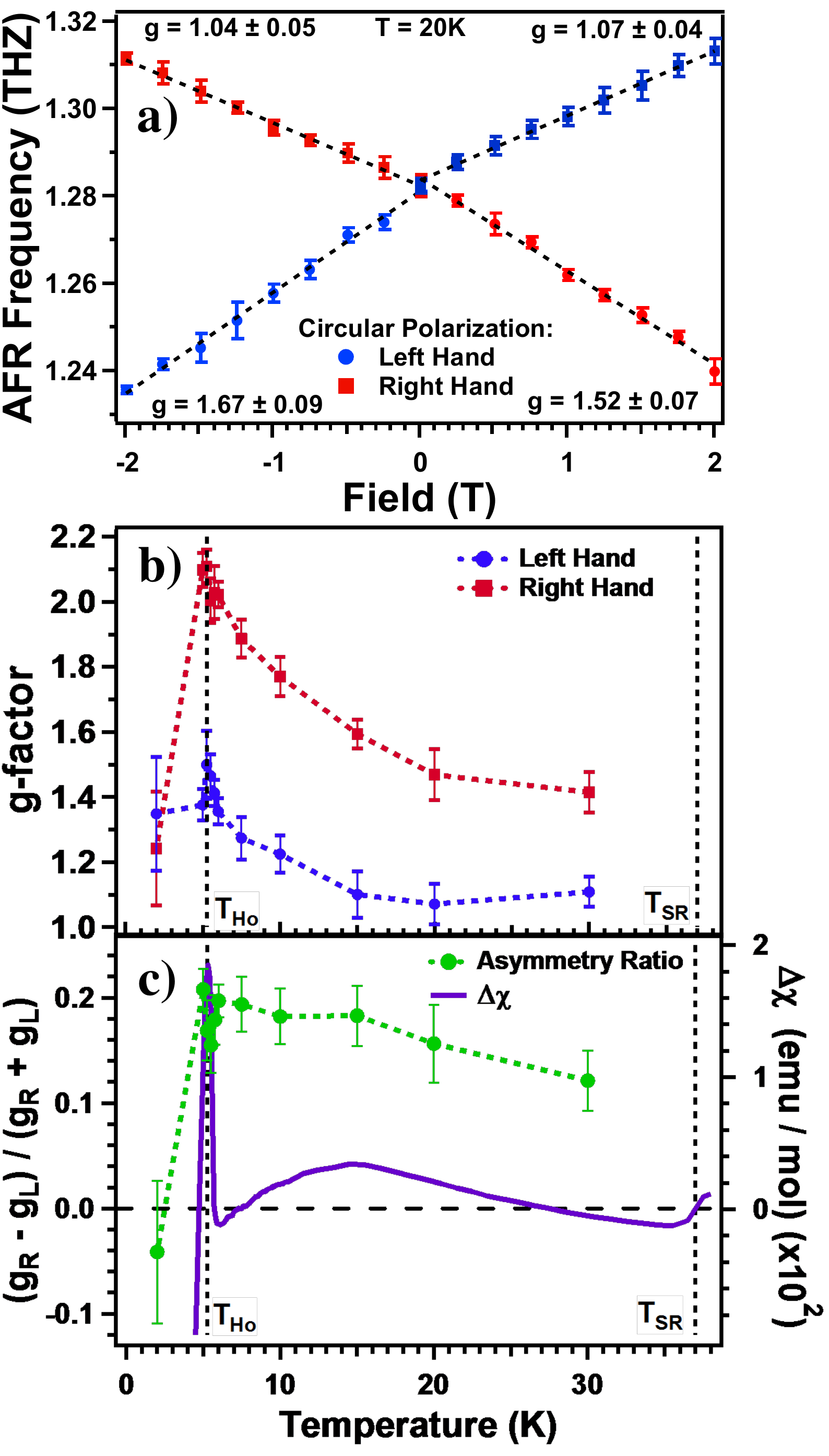}
\caption{(a) Resonant frequency of the AFR for both left hand (blue, circles) and right hand (red, squares) circular polarizations as a function of magnetic field at T = 20K.  The low energy branch of the AFR possesses a significantly larger $g$-factor than that of the high energy branch, regardless of polarization and field direction. (b) Temperature dependence of the $g$-factors which reveals a significant renormalization at T$_\text{Ho}$.  (c) Asymmetry ratio of the $g$-factors plotted with $\Delta \chi$, the $H$ $\parallel$ $c$ magnetic susceptibility after the paramagnetic contribution has been subtracted.}
\label{Fig3}
\end{figure}

The $g$-factors of the AFR can be found by fitting the extracted resonant frequencies as a function of magnetic field.  To reiterate, the expectation from \autoref{LowHAFR} is a symmetric splitting of the two branches with g-factors $\approx$ $\pm 1$. \autoref{Fig3}(a) displays linear fits of the AFR resonant frequencies in weak magnetic fields, within the low field phase of HMO.  One can see that $g$-factors are not only large but also unexpectedly \textit{asymmetric}, with the low energy branch possessing a $g$-factor that is $\approx$ 50\% greater than that of the high energy branch.  This asymmetry extends to negative fields as well, such that the low energy branch always possesses a larger $g$-factor.  A remarkable aspect of the data is the kink in the R and L branches as a function of B near zero field.  We believe this non-analyticity results from the manner in which the ground state is selected with a change in sign of the magnetic field as discussed below.  The small difference in $g$-factor for the low energy branches between positive and negative fields likely stems from larger error bars in negative fields due to a weak AFR in this orientation.  While enhanced $g$-factors have been interpreted via R-Mn spin interactions \cite{Talbayev2008}, asymmetry in the field dependent splitting of the AFR has not been reported previously.

We can ascertain the origin of this asymmetry by examining the temperature dependence of the $g$-factors (\autoref{Fig3}(b)).  The $g$-factors increase with decreasing temperature, a trend which is consistent with other hexagonal manganites \cite{Standard2012}.  However, In HMO we observe a large renormalization of the $g$-factors at T$_\text{Ho}$, with increases of $\approx$ 50\% and 35\% from 30K to T$_\text{Ho}$ in the right hand and left hand branches respectively.  This effect can be attributed to a large increase in the effective internal fields near T$_\text{Ho}$ as the Ho sublattices are more easily magnetized near the transition, consistent with the observed peak in the magnetic susceptibility at T$_\text{Ho}$ (Figure \ref{Fig3}(c)) \cite{Lorenz2013}. Below the transition, with the Ho sublattices presumably AF ordered, the internal fields are reduced and the $g$-factors return close to their high temperature values (although the errors bars at 2K are large due to overlap with neighboring Ho crystal field levels).  \autoref{Fig3}(c) displays the asymmetry ratio, defined as $(g_R - g_L) / (g_R + g_L)$, along with $\Delta \chi$, the $H$ $\parallel$ $c$ magnetic susceptibility of HMO after the paramagnetic contribution has been subtracted.  We can attribute this susceptibility to mainly stem from Ho magnetism.  One can see that the temperature dependence of the asymmetry ratio is in remarkable agreement with the magnetic susceptibility, increasing below T$_\text{SR}$, being renormalized at T$_\text{Ho}$, and decreasing rapidly at lower temperatures.  Such a plot unambiguously demonstrates the $g$-factor asymmetry to be related to Ho-Mn interactions.

To investigate the origin of this asymmetry, we have explored a scenario in which paramagnetic Ho moments generate an effective exchange field on Mn sites. To tilt both AFR branches down, this exchange field $\mathbf H_\mathrm{eff}$ must be antiparallel to Mn moments, reducing the cost of small deviations from the ordered state. At the same time, it is generated by fluctuating Ho moments whose thermal average $\langle S^z_\mathrm{Ho}\rangle = \chi_\mathrm{Ho} B$ is proportional to the applied field $B$ and to the Ho magnetic susceptibility $\chi_\mathrm{Ho}$, which grows as the temperature is lowered toward the Ho ordering at $\text{T}_\mathrm{Ho}$.  In this case, the exchange and anisotropy  of Eq. \ref{LowHAFR} are modified such that they depend linearly on $B$ as: 
\begin{equation}
J \mapsto J(B) \equiv J + J' B, 
\quad
\Delta \mapsto \Delta(B) \equiv \Delta + \Delta' B. 
\end{equation}
Then, to the linear order, the energies of the AFR are: 
\begin{equation}
\hbar \omega_\pm(B) = \hbar \omega(0) 
	+ \hbar \omega(0) 
		\left(
			\frac{J'}{2J} + \frac{\Delta'}{2\Delta} 
		\right)B 
	\pm g_{\text{eff}} \mu_B B
\end{equation}
and the slopes $d\omega_\pm/dB$ can differ in magnitude.
 
Such an exchange field in the $ab$ plane coming from Ho spins polarized along the $c$ axis can come from the Dzyaloshinskii-Moriya (DM) interaction, $H_\mathrm{DM} = \mathbf D \cdot (\mathbf S_\mathrm{Ho} \times \mathbf S_\mathrm{Mn})$, with a DM vector $\mathbf D$ in the ab plane \cite{Dz1958, Moriya1960}.  Although we believe this DM term plays a role here, in the most straightforward scenario this leads to the opposite effect: both AFR branches tilt up. To understand why, note that the effective exchange field $\mathbf H_\mathrm{eff} = - \mathbf D \times \langle \mathbf S_\mathrm{Ho}\rangle$ breaks the global symmetry of rotations in the $ab$ plane manifest in the Hamiltonian (\ref{MNSpinHam}). Mn spins orient themselves parallel to $\mathbf H_\mathrm{eff}$ to minimize the DM energy and select a ground state.  In general, it is this change in the ground state with the change in field direction that leads to the non-analyticity of the R and L excitations near B=0.  Deviations from these preferred directions now cost extra energy, which leads to a hardening of both AFR branches contrary to the experimental observations. We have found that other types of interactions breaking the global rotational symmetry that select a ground state generically harden both AFR branches \cite{SI}.  In order to get a softening, one must have the combined effect of both DM interaction and quartic interactions that force an anisotropy in-plane.  For instance, the interaction:
\begin{equation}
H_4 = K \sum_{\langle \mathrm{Ho} \mathrm{Mn} \rangle} 
	S_\mathrm{Ho}^z S_\mathrm{Mn}^{y}[3{(S_\mathrm{Mn}^{x})}^{2} - ({S_\mathrm{Mn}^{y})}^{2}],
\end{equation}
has been previously proposed to drive magnetic transitions in HMO \cite{Condran2010}.  However, other symmetry permitted quartic terms can also reproduce the observed asymmetry in the AFR (see Sec. III of the SI).  When both perturbations are present, one may select the ground state and the other determines the stiffness of the hard modes resulting in a net softening. This is a generic mechanism that may lead to g-factor asymmetry in other systems as well.

In summary, high precision time-domain THz experiments uncovered an asymmetric splitting of an AFR of the Mn sublattice in the multiferroic HMO.  Careful examination of the temperature dependence of this asymmetry unambiguously demonstrated the effect to be related to Ho-Mn interactions.  Theoretical analyses found this asymmetry is only reproduced if quartic spin interactions between Ho-Mn moments are included in the spin Hamiltonian. One generally expects such interactions to be present in other hexagonal manganites with rare-earth magnetism.  For instance, close inspection of the data of Ref. \cite{Standard2012} reveals that the low energy branch of the AFR possesses a significantly larger $g$-factor than the high energy branch in TmMnO$_3$, similar to our results in HMO.  Our analysis suggests that such interactions may be a general feature of exceptionally low symmetry antiferromagnets and warrant consideration.

Research at JHU was funded by the U.S. Department of Energy, Office of Basic Energy Sciences, Division of Materials Sciences and Engineering through Grant No. DE-FG02-08ER46544.  Work at Rutgers was supported by the DOE under Grant No. DOE: DE-FG02-07ER46382.  NJL acknowledges additional support through the ARCS Foundation Dillon Fellowship.  We would like to thank C. Broholm, M. Fiebig, R. Prasankumar, A. Sirenko, and D. Talbayev for helpful conversations.

\bibliography{HoMnO3_Mag}

\widetext

\section{Supplementary information for ``Asymmetric splitting of an antiferromagnetic resonance via quartic exchange interactions in multiferroic hexagonal HoMnO$_3$"}

\author{N. J. Laurita}
\affiliation{The Institute for Quantum Matter, Department of Physics and Astronomy, The Johns Hopkins University, Baltimore, MD 21218, USA}

\author{Yi Luo}
\affiliation{The Institute for Quantum Matter, Department of Physics and Astronomy, The Johns Hopkins University, Baltimore, MD 21218, USA}

\author{Rongwei Hu}
\affiliation{Rutgers Center For Emergent Materials, Department of Physics and Astronomy, Rutgers University, Piscataway, NJ, 08854, USA}

\author{Meixia Wu}
\affiliation{Rutgers Center For Emergent Materials, Department of Physics and Astronomy, Rutgers University, Piscataway, NJ, 08854, USA}

\author{S. W. Cheong}
\affiliation{Rutgers Center For Emergent Materials, Department of Physics and Astronomy, Rutgers University, Piscataway, NJ, 08854, USA}

\author{O. Tchernyshyov}
\affiliation{The Institute for Quantum Matter, Department of Physics and Astronomy, The Johns Hopkins University, Baltimore, MD 21218, USA}

\author{N. P. Armitage}
\affiliation{The Institute for Quantum Matter, Department of Physics and Astronomy, The Johns Hopkins University, Baltimore, MD 21218, USA}

\date{\today}

\maketitle

\section{Transmission Matrix In The Linear and Circular Bases} 

\begin{figure}[h]
\includegraphics[width=1.0\columnwidth,keepaspectratio]{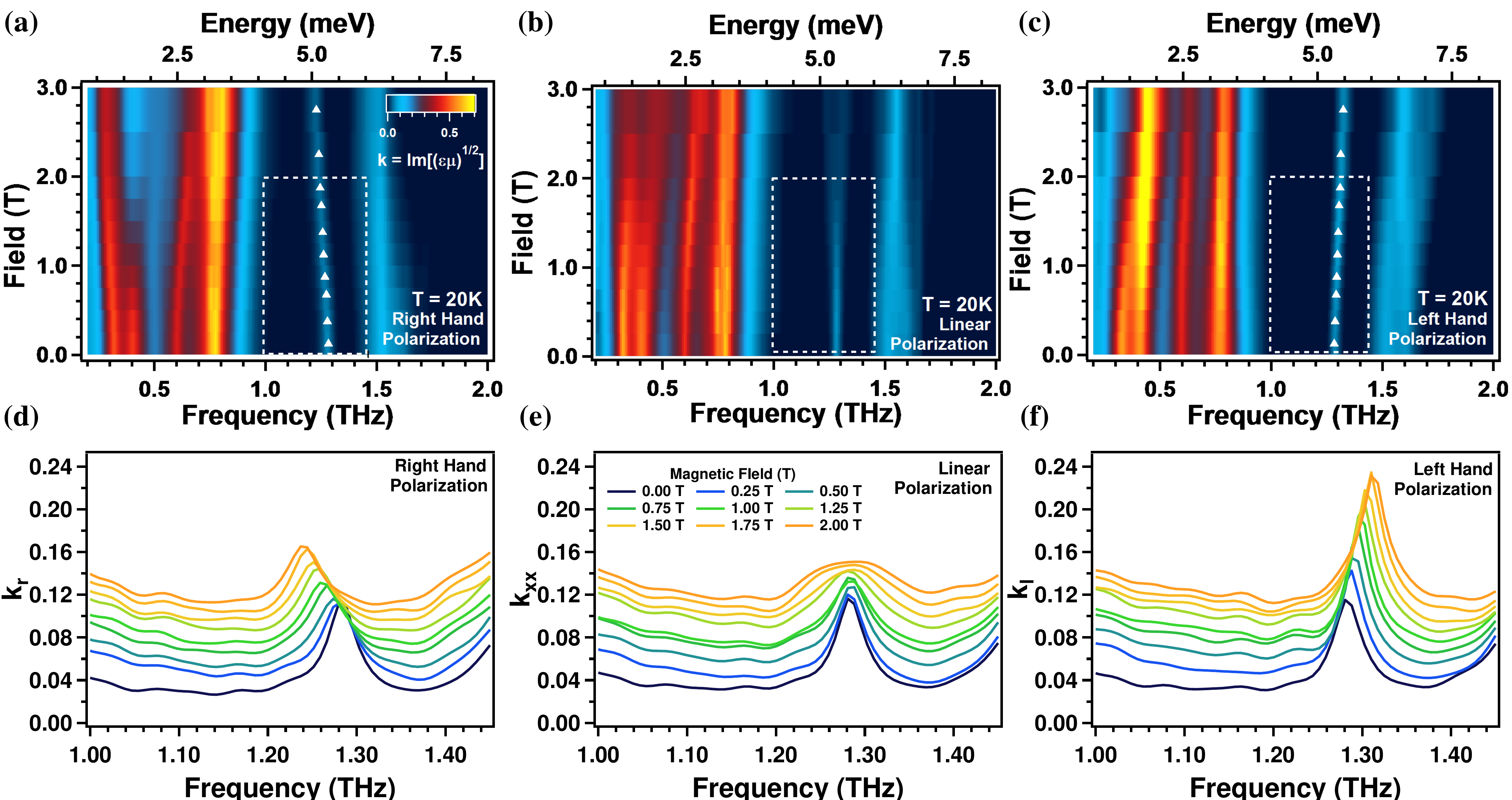}
\caption{Image plots of the dissipative part of the index of refraction in both the (b) linear basis and the (a) right and (c) left hand channels of the circular basis within the low field ``intermediate" phase of HMO.  (d)-(f) Plots of the dissipative part of the index of refraction within the dotted white box of the image plots shown in (a)-(c).  The advantage of converting to the circular basis is immediately apparent, as both modes of the AFR highly overlap in the linear basis but are naturally partitioned into the right and left hand channels of the circular basis.}
\label{SIFig2}
\end{figure}

In this section we detail the analysis of our time-domain terahertz (TDTS) data in both the linear and circular bases.  TDTS is a high resolution method for accurately measuring the electromagnetic response of a sample in the experimentally challenging THz range.  In a typical TDTS experiment, the electric field of a transmitted THz pulse through a sample is measured as a function of real time. Fourier transforming the measured electric field and referencing to an aperture of identical size allows access to the frequency dependent \textit{complex} transmission spectrum of the sample which, in the limit of $\tilde{\epsilon}$ $\gg$ $\tilde{\mu}$, is given by:
\begin{equation}
\widetilde{T} = \frac{4\widetilde{n}}{(\widetilde{n}+1)^2}\exp{{[\frac{i \omega d}{c}(\widetilde{n}-1)]}}
\label{Trans}
\end{equation}
where $d$ is the sample thickness, $\omega$ is the frequency, $c$ is the speed of light, $\widetilde{n}$ is the sample's complex index of refraction, and normal incidence has been assumed.  One can then numerically invert the transmission to obtain both the frequency dependent real and imaginary parts of the index of refraction.  In principle the index of refraction, $\widetilde{n} = n + ik = \sqrt{\tilde{\epsilon} \tilde{\mu}}$, contains both the electric and magnetic responses of the sample as THz fields can couple to both electric and magnetic dipole transitions.  Thus, electric and magnetic effects can be difficult to separate in such a single pass transmission experiment.  This is further complicated by the strong multiferroic and magnetoelectric coupling of HMO which introduces an additional term into the index of refraction.  Therefore, we neglect to attempt to identify features in the spectra as purely electric or magnetic in origin and instead report the combined response in the form of the imaginary, or dissipative, part of the index of refraction.

In the Jones calculus \cite{Jones1941}, the full linear response of the sample is represented as a complex 2 $\times$ 2 transmission matrix which transforms the incident electric field ($\mathbf{\tilde{E}_{\text{in}}}$) of THz light, written in the linear basis, as:
\begin{equation}
\begin{bmatrix}
\tilde{E}_\text{x,out} \\ \tilde{E}_\text{y,out}
\end{bmatrix}
=
\begin{bmatrix}
\tilde{T}_{xx} & \tilde{T}_{yx} \\ \tilde{T}_{xy} & \tilde{T}_{yy}
\end{bmatrix}
\begin{bmatrix}
\tilde{E}_\text{x,in} \\ \tilde{E}_\text{y,in}
\end{bmatrix}
\end{equation}
In the most general case the transmission matrix contains four independent components as shown above and one must vary the incident and detected polarization of light to measure individual elements of the transmission matrix.  However, Neumann's principle ensures that the transmission matrix must also possess the symmetries inherent to the crystal itself \cite{Armitage2014}, which often results in degeneracies between elements of the transmission matrix.  Such degeneracies can be exploited to measure samples in the basis in which the transmission matrix is diagonalized, and therefore the natural eigenpolarizations of the crystal.

TDTS experiments performed as a function of magnetic field were done so in the Faraday geometry, in which $\vec{\text{k}}_\text{THz}$ $\parallel$ $\vec{\text{H}}_{\text{dc}}$ $\parallel \hat{z}$, on a HMO sample oriented such that $c$ $\parallel \hat{z}$.  In this case, the crystal structure is symmetric under rotations of 120$^\circ$ about the $\hat{z}$ axis.  The transmission matrix must therefore also obey this symmetry, which greatly constrains its form such that it must be fully antisymmetric with only two unique components: 
\begin{equation}
\tilde{T}_\text{linear}
=
\begin{bmatrix}
\tilde{T}_{xx} & \tilde{T}_{xy} \\ -\tilde{T}_{xy} & \tilde{T}_{xx}
\end{bmatrix}
\end{equation}
Such a fully antisymmetric transmission matrix can be diagonalized by a circular basis transformation as: 
\begin{equation}
\tilde{T}_\text{circular}
=
\begin{bmatrix}
\tilde{T}_{xx} + i \cdot \tilde{T}_{xy} & 0 \\ 0 & \tilde{T}_{xx} - i \cdot \tilde{T}_{xy}
\end{bmatrix}
=
\begin{bmatrix} \tilde{T}_{r}&0 \\ 0 & \tilde{T}_{l} \end{bmatrix}
\end{equation}
where $\tilde{T}_l$ and $\tilde{T}_r$ refer to the transmission of left and right hand circularly polarized light, the eigenpolarizations for this orientation, respectively.  The above analysis suggests that experiments performed in Faraday geometry are best understood in the circular basis.

\begin{figure}[t]
\includegraphics[width=1.0\columnwidth, keepaspectratio]{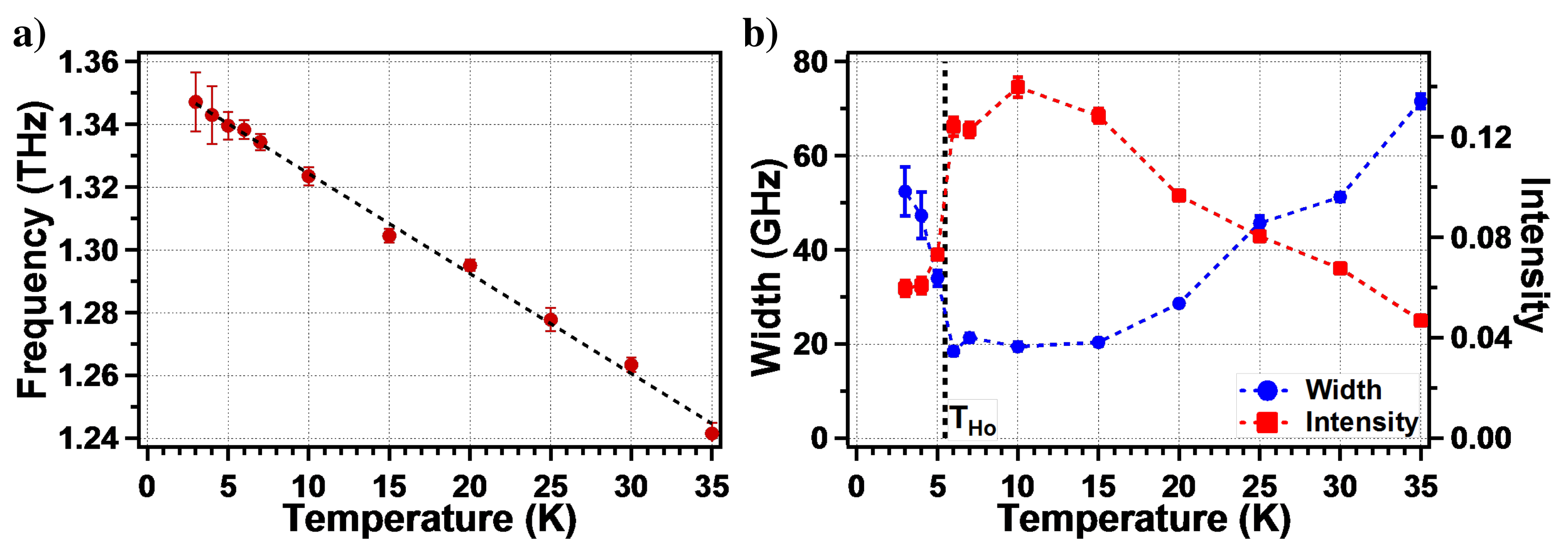}
\caption{Temperature dependent dynamical properties of the Mn AFR for temperatures below T$_\text{SR}$.  (left) Resonant frequency. (right) Width (blue squares, left axis) and intensity (red circles, right axis).}
\label{SIFig10}
\end{figure}
  
As we demonstrated in the main text such a conversion to the circular basis can be highly advantageous in the study of magnetic excitations.  \autoref{SIFig2} displays magnetic field dependent data in both the linear and circular basis to demonstrate the advantage of such a transformation.  Shown are image plots of the imaginary part of the index of refraction in the (b) linear and (a) right hand and (c) left hand circular polarizations.  Plots (d)-(f) display the data in the regions marked by the white dotted boxes of (a)-(c), the range fit to extract the $g$-factors in Figure 3(a) of the main text.  One can see the difficulty in extracting the resonant frequencies of both branches of the AFR in the linear basis, as the field dependent splitting is far smaller than the excitation width such that the two branches of the AFR highly overlap, even at fields as large as a few Tesla. With such a complex magnetic field dependent phase diagram as HMO \cite{Vajk2005}, the study of the AFR in the low field phase has been previously impossible.  However, the high and low energy branches of the AFR naturally partition into left and right hand circular polarizations respectively in the circular basis, allowing for determination of their resonant frequencies and $g$-factors with exceptional precision in weak magnetic fields.  We expect such a transformation to be similarly beneficial to the study of other hexagonal manganites as well, thus providing a paradigm for the optical study of the AFR in these and other classes of materials.

\section{Temperature and Field Dependence of the Mn AFR}

The temperature dependence of the resonant frequency of the Mn AFR has been previously investigated in several  hexagonal manganites \cite{Standard2012}.  In these systems it was empirically found that the AFR frequency displays a power law dependence with temperature as $\omega(T)$ = $\omega(0)$ + $aT^b$, where the exponent $b$ was found to be $\approx$ 3.  Deviations from this power law behavior at low temperatures were observed in hexagonal manganites with rare-earth magnetism and were interpreted as corresponding to the onset of rare-earth moment fluctuations and therefore RE-Mn interactions. 

\begin{figure}[t]
\includegraphics[width=1.0\columnwidth, keepaspectratio]{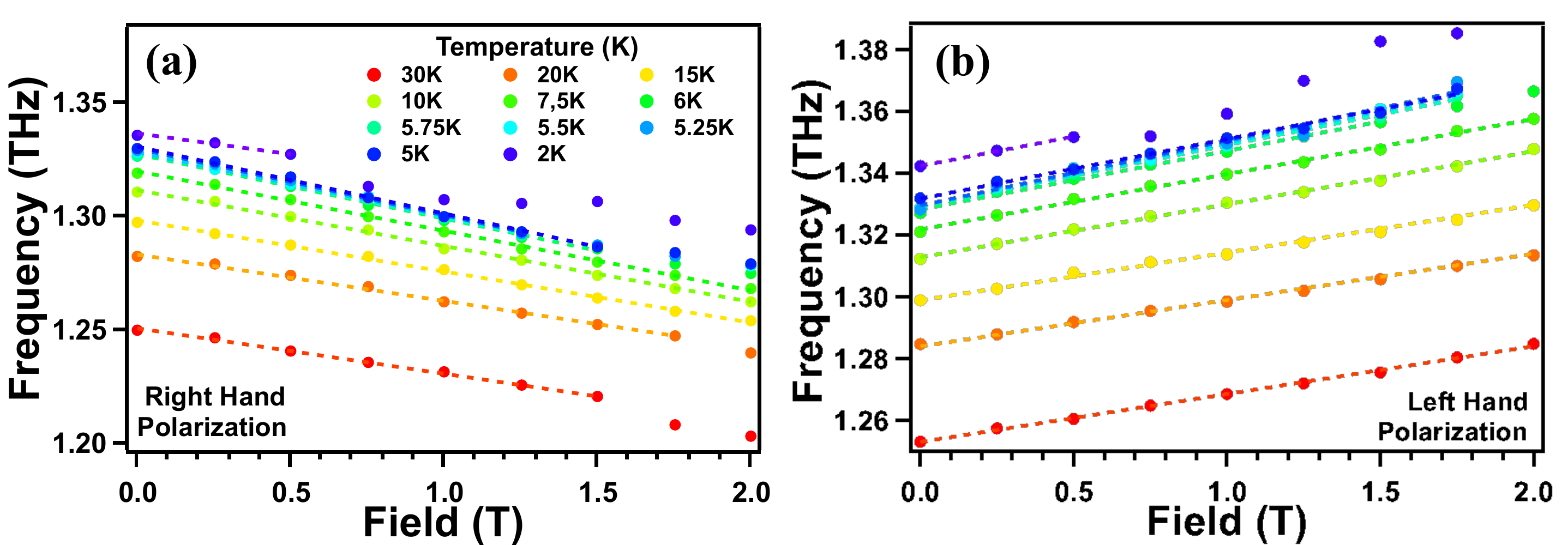}
\caption{Field dependence of the Mn AFR resonant frequency in the right and left hand channels of the circular basis.  Data is shown below 2T, where the highest signal to noise is achieved and well within the low field phase of HMO.  Dotted lines are the resultant linear fit of the data at each temperature.}
\label{SIFig11}
\end{figure}

\autoref{SIFig10}(a) displays the resonant frequency of the AFR in HMO for temperatures below T$_{\text{SR}}$, extracted by fitting the spectra as described in the main text.  One can see that in contrast to other hexagonal manganites, the AFR frequency in HMO displays a linear dependence ($b$ = 1.02) with temperature.  This is consistent with the onset of magnetic fluctuations of at least one of the Ho sublattices at T$_\text{SR}$, in agreement with previous reports \cite{Lonkai2002, Nandi2008} and our own analysis presented in the main text.  \autoref{SIFig10}(b) displays the temperature dependence of the width (blue, left axis) and intensity (red, right axis) of the AFR for temperatures below T$_\text{SR}$.  One can see that as the temperature is reduced the AFR becomes more well defined, displaying both an increase in intensity and narrowing.  However, at T$_\text{Ho}$ the AFR is dramatically damped and the intensity is significantly reduced.  The origin of such an effect is currently unclear but one possible explanation is that the Mn AFR hybridizes with nearly degenerate Ho crystal field levels below T$_\text{Ho}$.  If so, then HMO may display the first hybridized magnon-crystal field excitations which originate from distinct magnetic sublattices.  Further research is required to confirm such a possibility.

\autoref{SIFig11} displays the AFR resonant frequency as a function of magnetic field for temperatures below T$_\text{SR}$ for both the (a) right hand and (b) left hand circular polarizations.  Markers denote the extracting resonant frequency found by fitting the data described in the text while dotted lines represent linear fits of the data from which the $g$-factors were extracted.  It was previously found that the $g$-factors of the Mn AFR change significantly in the ``high field" phase of hexagonal manganites, i.e. for field H $>$ H$_c$ $\approx$ 3 T  \cite{Standard2012}.  Therefore, care was taken to only include data taken within the low field ``intermediate" phase to obtain the $g$-factors.  Such phase boundaries were identified by obvious non-linearities in our data as well as published phase boundaries found previously in HMO \cite{Vajk2005}.   

\section{Spin waves in an easy-plane triangular antiferromagnet}
\subsection{Spin Lagrangian}

We first derive the frequency of the antiferromagnetic resonance in an easy-plane triangular antiferromagnet in the presence of a weak magnetic field normal to the easy plane. The spin Hamiltonian is 
\begin{equation}
H = J \sum _{\langle ij \rangle} \mathbf S_i \cdot \mathbf S_j 
	+ \Delta \sum_{i} ({S_{i}^{z})}^2 
	+ g \mu_B \mathbf B \cdot \sum _{i} \mathbf S_{i},
\label{MNSpinHam}
\end{equation}
Here $J>0$ is the nearest-neighbor exchange coupling, $\Delta>0$ is the strength of easy-plane anisotropy, $g$ is the Land{\'e} gyromagnetic factor, $\mu_B$ is the Bohr magneton, and $\mathbf B = (0,0,B)$ is an external magnetic field. The index $i$ is summed over all spins, $\langle ij \rangle$ over all nearest-neighbor pairs of spins. We treat the spins as classical vectors of fixed length $\hbar S$ and parametrize them in terms of spherical angles $\theta_i$ and $\phi_i$ in the global frame $xyz$: 
\begin{equation}
\mathbf S_i = S(\sin{\theta_i} \cos{\phi_i}, \sin{\theta_i} \sin{\phi_i}, \cos{\theta_i}).
\end{equation}

\begin{figure}[ht]
\includegraphics[width=0.5\columnwidth]{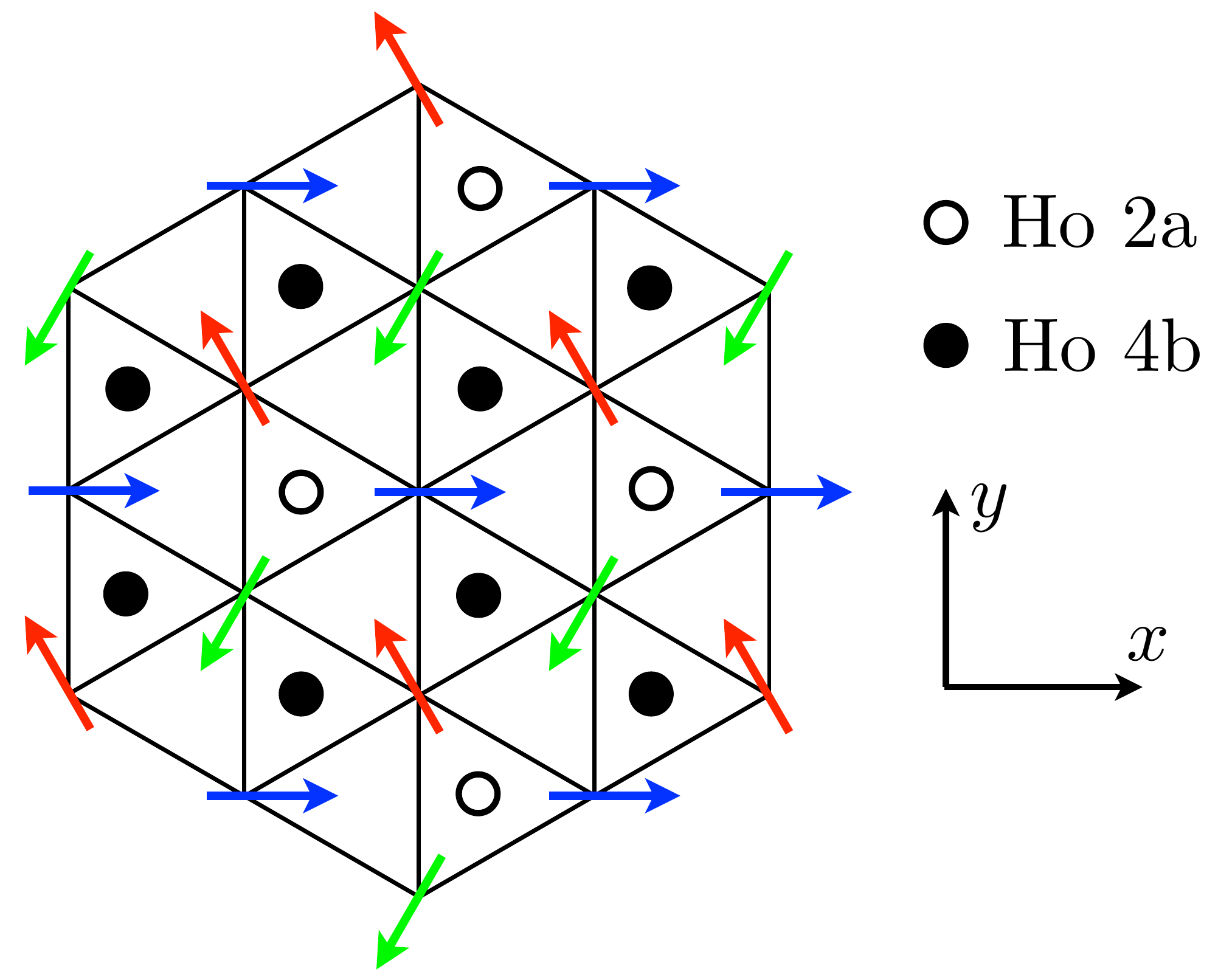}
\caption{The 3-sublattice antiferromagnetic order on the Mn lattice (red, green, and blue arrows) and the two types of Ho ions, 2a (open circles) and 4b (filled circles).}
\label{fig:order}
\end{figure}

In equilibrium, the spins form an ordered pattern with three sublattices $n = 1, 2, 3$ (Fig.~\ref{fig:order}): 
\begin{equation}
\cos{\theta_n} = \frac{g \mu_B B}{(9J + 2\Delta)S},
\quad
\phi_n = \phi_z + \frac{2\pi n}{3},
\end{equation}
A spin on one sublattice has three neighbors from each of the other sublattices. The offset angle $\phi_z$ represents the global symmetry of rotations about the hard axis $z$. 

\begin{figure}[ht]
\includegraphics[width=0.8\columnwidth]{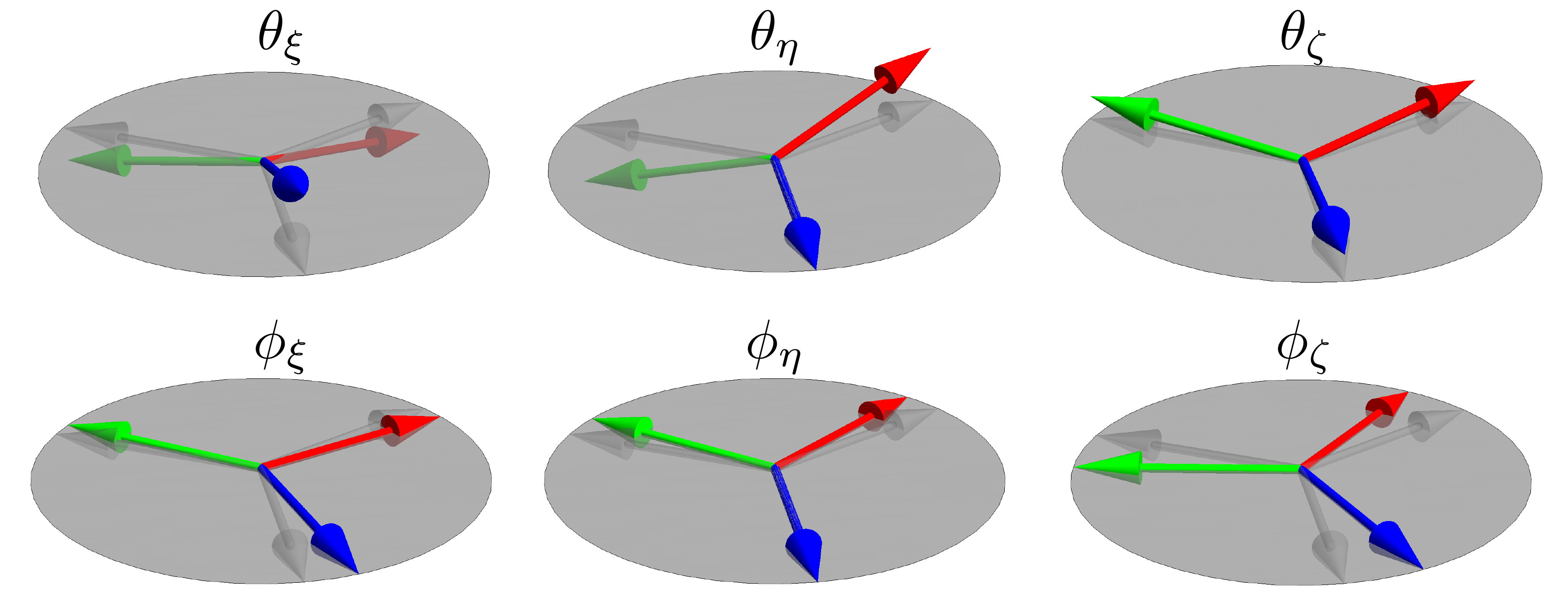}
\caption{Normal modes of a triangular antiferromagnet. In zero field, the three sublattice spins $\mathbf S_1$ (red), $\mathbf S_2$ (green), and $\mathbf S_3$ (blue) lie in the easy plane $xy$ (gray). Faint arrows represent an equilibrium state with the spins in the easy plane and pointing at $120^\circ$ to one another. The $\theta$ and $\phi$ modes move the spins in the polar and azimuthal directions, respectively. The spin axes $\xi$ and $\eta$ point along $\mathbf S_2 - \mathbf S_1$ and $\mathbf S_3$, respectively. The $\zeta$ axis coincides with the hard axis $z$.}
\label{fig:modes}
\end{figure}

The dynamics of small-amplitude spin waves with $\mathbf q = 0$ is most readily obtained from the Lagrangian 
\begin{equation}
L = \hbar S \sum_{n=1}^3 (\cos{\theta_n}-1) \dot{\phi}_n - U(\{\theta,\phi\}).
\end{equation}
The first term in the Lagrangian represents the geometric (Berry-phase) part of the spin action, the second is potential energy. 

\subsection{Spin waves}

For small deviations from equilibrium, we introduce normal coordinates representing eigenmodes: 
\begin{equation}
\left(
	\begin{array}{c}
		\delta \theta_1 \\
		\delta \theta_2 \\
		\delta \theta_3
	\end{array}
\right)
 = 
\left(
	\begin{array}{rrr}
		 \frac{1}{2} & -\frac{\sqrt{3}}{2} & -\frac{1}{3} \\
		\frac{1}{2} & \frac{\sqrt{3}}{2} & -\frac{1}{3} \\
		-1 & 0 & -\frac{1}{3}
	\end{array}
\right)
\left(
	\begin{array}{c}
		\theta_\xi \\
		\theta_\eta \\
		\theta_\zeta
	\end{array}
\right),
\qquad
\left(
	\begin{array}{c}
		\delta \phi_1 \\
		\delta \phi_2 \\
		\delta \phi_3
	\end{array}
\right)
 = 
\left(
	\begin{array}{rrr}
		-\frac{1}{3} & \frac{1}{\sqrt{3}} & 1 \\
		-\frac{1}{3} & -\frac{1}{\sqrt{3}} & 1 \\
		\frac{2}{3} & 0 & 1
	\end{array}
\right)
\left(
	\begin{array}{c}
		\phi_\xi \\
		\phi_\eta \\
		\phi_\zeta
	\end{array}
\right)
\end{equation}
For convenience, we have introduced a new frame $\xi\eta\zeta$ defined by the ordered state (Fig.~\ref{fig:modes}). The normal modes have the following meanings in the absence of the applied field. $\theta_\zeta$ describes the average latitude of the three sublattice spins; thus $\hbar S \theta_\zeta$ is the net spin in the $\zeta$ direction, a momentum conjugate to the global rotation angle $\phi_\zeta$ in the easy plane. (The mode $\phi_\zeta$ is redundant as it globally rotates all spins about the hard axis, thus playing the same role as $\phi_z$ does.) Global rotations tilting the spins out of the easy plane are quantified by angles $\theta_\xi$ and $\theta_\eta$ about the respective spin axes. Lastly, $\phi_\xi$ and $\phi_\eta$ represent deformations of the $120^\circ$ order within the easy plane and create a net spin with projections $-\hbar S \phi_\xi$ and $-\hbar S \phi_\eta$, onto the $\xi$ and $\eta$ axes respectively. 

To the second order in the normal coordinates and to the first order in the magnetic field $B$,
\begin{equation}
L = \hbar S 
	(\theta_\xi \dot{\phi}_\xi + \theta_\eta \dot{\phi}_\eta + \theta_\zeta \dot{\phi}_\zeta)
	- U,
\end{equation}
where 
\begin{equation}
U = \frac{3 J S^2}{2} 
		\left(
			\phi_\xi^2 + \phi_\eta^2
		\right)
	+ \frac{3 \Delta S^2}{2} 
		\left(
			\theta_\xi^2 + \theta_\eta^2
		\right)
	+ \tilde{g} \mu_B B S
		\left(
			\theta_\xi \phi_\eta - \theta_\eta \phi_\xi
		\right)
	+ \frac{(9J + 2\Delta)S^2}{6} \theta_\zeta^2
\label{eq:U}
\end{equation}
is the potential energy and 
\begin{equation}
\tilde{g} = \frac{g}{2(1+2\Delta/9J)}
\end{equation}
is an effective Land{\'e} gyromagnetic factor.

As expected, the soft mode $\phi_\zeta$ does not influence the potential energy and its conjugate momentum $\hbar S \theta_\zeta$ is a conserved quantity. These two coordinates represent the Goldstone mode with frequency $\omega = 0$ in the limit $\mathbf q \to 0$. 

The equations of motion for the remaining four coordinates are
\begin{equation}
\hbar \frac{d}{dt}
	\left(
		\begin{array}{c}
			\theta_\xi \\
			\phi_\xi \\
			\theta_\eta \\
			\phi_\eta
		\end{array}
	\right) 
= 
	\left(
		\begin{array}{cccc}
			0 & - 3JS & \tilde{g} \mu_B B & 0 \\
			3 \Delta S & 0 & 0 & \tilde{g} \mu_B B \\
			- \tilde{g} \mu_B B & 0 & 0 & - 3JS \\
			0 & - \tilde{g} \mu_B B & 3 \Delta S & 0
		\end{array}
	\right)
	\left(
		\begin{array}{c}
			\theta_\xi \\
			\phi_\xi \\
			\theta_\eta \\
			\phi_\eta
		\end{array}
	\right)
\end{equation}
In zero magnetic field, there are two degenerate magnons with energy $\hbar \omega(0) = 3S\sqrt{J\Delta}$, one involving $\theta_\xi$ and $\phi_\xi$, the other $\theta_\eta$ and $\phi_\eta$. The field couples the two modes, lifting the degeneracy. The eigenmodes and eigenfrequencies are 
\begin{equation}
\left(
	\begin{array}{c}
		\theta_\xi \\
		\phi_\xi \\
		\theta_\eta \\
		\phi_\eta
	\end{array}
\right) 
= 
\left(
	\begin{array}{l}
		\sqrt{J} \cos{\omega t} \\
		\sqrt{\Delta} \sin{\omega t} \\
		\sqrt{J} \cos{(\omega t \pm \pi/2)} \\
		\sqrt{\Delta} \sin{(\omega t \pm \pi/2)}
	\end{array}
\right),
\qquad
\hbar \omega_\pm(B) = 3S\sqrt{J\Delta} \pm \tilde{g} \mu_B B.
\end{equation}
The in-plane components of spin, $\hbar \mathbf S_\perp = -\hbar S (\phi_\xi, \phi_\eta, 0)$ in the $\xi\eta\zeta$ frame, rotate clockwise or counterclockwise, generating circularly polarized electromagnetic waves. The straight lines $\omega_\pm(B)$ have equal and opposite slopes $d\omega_\pm/dB = \pm \tilde{g} \mu_B/\hbar$.

Different slope magnitudes will result if we modify the first two terms in the potential energy (\ref{eq:U}) so that their coefficients depend linearly on $B$. We may replace 
\begin{equation}
J \mapsto J(B) \equiv J + J' B, 
\quad
\Delta \mapsto \Delta(B) \equiv \Delta + \Delta' B. 
\end{equation}
Then, to the linear order in $B$, the eigenfrequencies will be 
\begin{equation}
\omega_\pm(B) = \omega(0) 
	+ \omega(0) 
		\left(
			\frac{J'}{2J} + \frac{\Delta'}{2\Delta} 
		\right)B 
	\pm \tilde{g} \mu_B B
\end{equation}
and the slopes $d\omega_\pm/dB$ will differ in magnitude.

\subsection{Dzyaloshinskii-Moriya interaction between Ho and Mn spins}

One of the mechanisms that could create a difference in the slopes $d\omega_\pm/dB$ is the Dzyaloshinskii-Moriya (DM) interaction $\mathbf D \cdot (\mathbf S_\mathrm{Ho} \times \mathbf S_\mathrm{Mn})$ between Ho and Mn spins. At intermediate temperatures, Ho spins remain disordered but are magnetized by the applied magnetic field, $\mathbf s = \chi \mathbf B$, where $\chi$ is the Ho susceptibility. With the Ho spins magnetized along the applied field (along the hard axis $z$), we only need to consider the transverse $x$ and $y$ components of the DM vector $\mathbf D$. They generate an effective magnetic field on an Mn site $\mathbf h_\mathrm{DM} = - \mathbf D \times \langle \mathbf S_\mathrm{Ho} \rangle$ oriented in the easy plane $xy$. For a given Mn site, the fields from the adjacent Ho ions vary in orientation. 

If all six adjacent Ho ions (three above and three below an Mn plane) were equivalent and located symmetrically, the Mn site would possess a threefold rotational symmetry and the net $\mathbf h_\mathrm{DM}$ would vanish. However, a lattice distortion breaks the symmetry and creates inequivalent Ho sites known as 2a and 4b. Each Mn has two 2a and four 4b Ho neighbors. As a result, the effective field $\mathbf h_\mathrm{DM}$ can be nonzero. Its orientation is restricted to the vertical plane containing the Mn ion and both of its 2a Ho neighbors, while its magnitude $h_\mathrm{DM} = C_\mathrm{DM} B$, where $C_\mathrm{DM}$ is a constant. 

The in-plane effective field $\mathbf h_\mathrm{DM}$ breaks the global rotational invariance with respect to $z$-axis rotations and selects a unique ground state, in which each Mn spin points along its local effective field $\mathbf h_\mathrm{DM}$. It adds the following term to the spin energy, to second order in the normal modes of interest: 
\begin{equation}
U_\mathrm{DM} 
	= C_\mathrm{DM} B \cos{\phi_z} 
		\left[
			-3 + \frac{3}{4} (\theta_\xi^2 + \theta_\eta^2) + \frac{1}{3} (\phi_\xi^2 + \phi_\eta^2)
		\right].
\label{UDM-five}
\end{equation}

Depending on the sign of the effective field $h_\mathrm{DM} = C_\mathrm{DM} B$, it selects the ground state with the global rotation angle $\phi_z = 0$ (if $C_\mathrm{DM} B > 0$) or $\phi_z = \pi$ (if $C_\mathrm{DM} B < 0$). The four modes stiffen: 
\begin{equation}
U_\mathrm{DM} 
	= |C_\mathrm{DM} B|  
		\left[
			-3 + \frac{3}{4} (\theta_\xi^2 + \theta_\eta^2) + \frac{1}{3} (\phi_\xi^2 + \phi_\eta^2)
		\right].
\end{equation}
Thus the presence of the DM interaction alone does not explain the softening of the two resonance branches $\omega_\pm(B)$.

\subsection{Quartic spin interactions}

Another route to an asymmetric splitting of the resonance is offered by higher-order spin interactions proposed by \textcite{condran2010model} that couples the out-of-plane component of Ho spins to the in-plane components of Mn spins:
\begin{equation}
U_4 = K \sum_{\langle \mathrm{Ho} \mathrm{Mn} \rangle} 
	S_\mathrm{Ho}^z \prod_{n=1}^3 (\mathbf S_\mathrm{Mn} \cdot \mathbf e_n),
\end{equation}
with the unit vectors
\begin{equation}
\mathbf e_n = \left(\cos{\frac{2\pi n}{3}}, \sin{\frac{2\pi n}{3}}, 0\right)
\end{equation}
in the global frame $xyz$. This quartic spin interaction, introduced on phenomenological grounds, is compatible the time reversal and $D_{3h}$ point-group symmetries. (Here we neglect the distortion creating inequivalent 2a and 4b Ho sites.) 

In an applied magnetic field $\mathbf B = (0,0,B)$, the energy of the modes of interest has the following additions:
\begin{equation}
U_4 
	= C_4 B \cos{3\phi_z}
		\left[
			- \frac{3}{4} + \frac{9}{16} (\theta_\xi^2 + \theta_\eta^2) + \frac{3}{4} (\phi_\xi^2 + \phi_\eta^2)
		\right].
\label{U4-five}
\end{equation}
The Mn spins now have three possible ground states. Depending on the sign of the prefactor $C_4 B$, the energy is minimized by the global azimuthal angles $\phi_z = 0, \pm \frac{2\pi}{3}$ (if $C_4 B > 0$) or $\phi_z = \pi, \pm \frac{\pi}{3}$ (if $C_4 B < 0$). In both cases, the AFR modes stiffen: 
\begin{equation}
U_4 
	= |C_4 B|
		\left[
			- \frac{3}{4} + \frac{9}{16} (\theta_\xi^2 + \theta_\eta^2) + \frac{3}{4} (\phi_\xi^2 + \phi_\eta^2)
		\right].
\end{equation}

\subsection{Combined interactions}

Although neither interaction alone is capable of reproducing the experimentally observed softening of both AFR branches, their combination can, at least in principle. When both perturbations are present, one of them may select the ground state and the other determines the stiffness of the hard modes. This is possible because the DM energy (\ref{UDM-five}) has a relatively large zeroth-order term, whereas the quartic interaction (\ref{U4-five}) has larger quadratic terms. 

For example, let us take $C_\mathrm{DM} > 0$ and $C_4 = - 2 C_\mathrm{DM} < 0$. The DM term dominates in selecting the ground state with $\phi_z = 0$ (for $B>0$). The energy of the AFR modes is
\begin{equation}
U_\mathrm{DM} + U_4 
	= |C_\mathrm{DM} B|
		\left[
			- \frac{3}{2} - \frac{3}{8}(\theta_\xi^2 + \theta_\eta^2) - \frac{7}{6} (\phi_\xi^2 + \phi_\eta^2)
		\right].
\end{equation}
The quadratic terms are expressly negative, so both AFR branches soften. 

Other interactions may be combined for a similar effect. For example, a quartic DM term $\mathbf D' \cdot (\mathbf S_\mathrm{Ho} \times \mathbf S_\mathrm{Mn}) S_\mathrm{Mn}^2$ would create a potential
\begin{equation}
U 
= -\frac{3}{2} C'_\mathrm{DM}B \cos{\phi_z} (\theta_\xi^2 + \theta_\eta^2).
\end{equation}

\section{Description Of The Crystal Field Spectrum}

\begin{figure}[t]
\includegraphics[width=0.9\columnwidth,keepaspectratio]{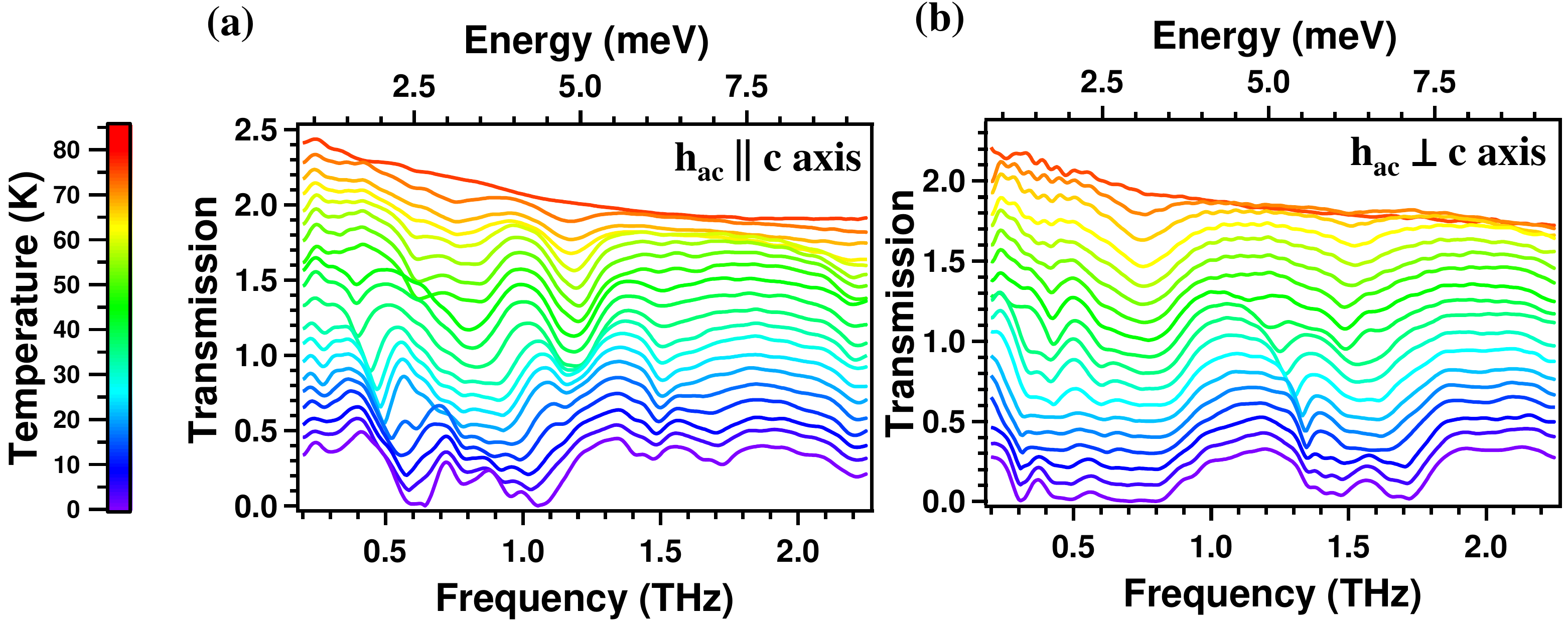}
\caption{(a)-(b) Magnitude of the complex transmission as a function of frequency and temperature for the (a) $\vec{\text{h}}_{\text{ac}}$ $\parallel$ $c$ and (b) $\vec{\text{h}}_{\text{ac}}$ $\perp$ $c$ geometries respectively.  Curves are offset by 0.1 per temperature for clarity.}
\label{SIFig3}
\end{figure}

\subsection{Crystal Field Theory}

We can begin to understand the crystal field spectrum of the Ho$^{+3}$ ions by examining the relevant energy scales sequentially.  One expects spin orbit coupling to be the dominant energy scale in 4$f$ electron systems due to the high atomic number of rare earth ions.  Assuming pure LS-coupling, the ground state of the Ho$^{+3}$ ions are determined by Hund's rules to be $^5$I$_8$.  With only spin orbit coupling considered, all the states within the J=8 manifold are degenerate and well separated from the nearest J=7 manifold of states.  The next largest expected energy scale is that imposed by the surrounding crystal field.  However, the energy scale of the crystal field is expected to be two orders of magnitude weaker than that of spin-orbit coupling, and therefore acts as a perturbation on the J=8 manifold.  In the point charge limit, the crystal field breaks the degeneracy of the J=8 manifold by coupling eigenstates of the $\hat{J}_z$ operator, depending directly on the symmetry of the crystal field at the rare earth site.  

As described above, the 2a and 4b Ho$^{+3}$ sites of the lattice are symmetry distinct, possessing C$_{3v}$ and C$_3$ point group symmetries respectively.  The effects of such crystal fields on the $^5$I$_8$ ground state of Ho$^{+3}$ have previously been studied in a number of compounds \cite{Judd1955, Abragam1970, Elliott1953, Rosenkranz2000, Ranon1969}.  The crystal field potentials, V$_{\text{CF}}$, for both symmetries are easily understood in Stevens notation \cite{Stevens1952} as:
\begin{gather}
V_{c_3} = B_0^2O_0^2 + B_0^4O_0^4 + B_0^6O_0^6 + B_6^6O_6^6 \\
V_{c_{3v}} = B_0^2O_0^2 + B_0^4O_0^4 + B_0^6O_0^6 + B_6^6O_6^6 + B_4^3O_4^3 + B_6^3O_6^3
\end{gather}
respectively.  Where the $B_l^q$ are the crystal field parameters and the $O_l^q$ are Stevens operators.  Stevens operators in the form $O_0^q$ form the diagonal of V$_\text{CF}$ and only depend on $\hat{J}$ and $\hat{J}_z$ and therefore do not couple states with differing $m_z$.  Operators in the form $O_l^q$ form the off-diagonal components of V$_\text{CF}$ and contain terms with $\hat{J}_{\pm}$, thereby coupling terms with $m_z$ that differ by $\pm q$.  Thus, for C$_{3v}$ symmetry states with $\ket{m_z}$ differing by either $\ket{\pm3}$ or $\ket{\pm6}$ are coupled.  The result is a spectrum of singlets and doublets in the form: 
\begin{gather}
\ket{\psi} = a\ket{6} + b\ket{3} + c\ket{0} + d\ket{-3} + e\ket{-6}\\ 
\ket{\phi} = a\ket{\pm{8}} + b\ket{\pm{5}} + c\ket{\pm{2}} + d\ket{\mp{1}} + e\ket{\mp{4}} + f\ket{\mp{7}}. 
\end{gather}
respectively.  Similarly, the C$_{3}$ crystal field couples states with $\ket{m_z}$ differing by $\ket{\pm6}$.  The result is still a spectrum of singlets and doublets with the doublet states being identical to those of the C$_{3v}$ crystal field listed above.  However, the singlets can take the form of either:
\begin{gather}
\ket{\phi} = a\ket{6} + b\ket{0} + c\ket{-6}\\
\ket{\phi} = \frac{1}{\sqrt{2}}(\ket{3} \pm \ket{-3}). 
\end{gather}  

The above calculation predicts a spectrum of singlets and doublets but the ordering of the energy levels is  generally determined experimentally by probing degeneracies via Zeeman interaction under applied field as well as the optical selection rules of the system.  With the quantization axis along the hexagonal $c$ direction, one expects that for h$\perp$c polarization to couple states with $\Delta$m$_z$ $=$ $\pm$1.  Thereby allowing transitions between singlets and doublets or between different doublets.  With h$\parallel$c, one expects transitions with $\Delta$m$_z$ $=$ 0 i.e. within doublets.  

However, additional effects which depart from the simplistic picture of only spin-orbit coupling and crystal field splitting outlined above must also be considered.  For instance, broken inversion symmetry will permits electric dipole excitations while  interactions the nuclear spin of Ho$^{+3}$ ions will permit hyperfine magnetic dipole and electric quadrupole terms into the Hamiltonian.  Magnetic exchange represents an additional energy scale which will further modify the spectrum.  All such effects will work in tandem to produce the crystal field spectrum observed in HMO.  It should noted that the J=8 Ho$^{+3}$ ions are non-Kramers ions and thus any degeneracy is not protected by symmetry.  Therefore, one may generally expect a crystal field spectrum which is considerably more complex than one of temperature independent singlets and doublets. 

\subsection{Temperature Dependence Of The Crystal Field Spectra}

\begin{figure}[t]
\includegraphics[width=0.85\columnwidth, keepaspectratio]
{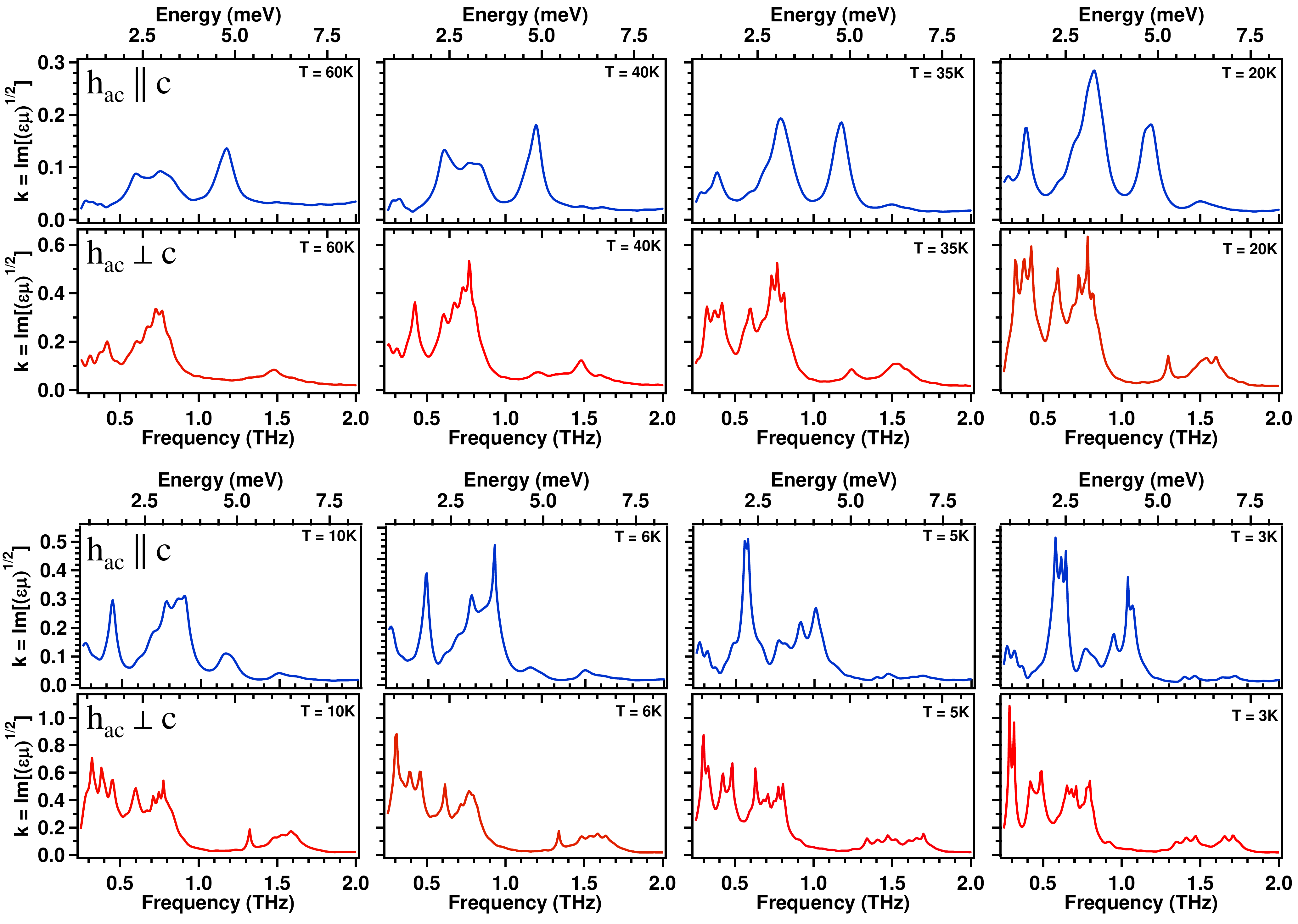}
\caption{The dissipative part of the index of refraction, k = $\operatorname{\mathbb{I}m}\{\sqrt{\epsilon \mu}\}$, for the $\vec{\text{h}}_{\text{ac}}$ $\parallel$ c (top rows, blue) and $\vec{\text{h}}_{\text{ac}}$ $\perp$ c orientations (bottom rows, red) at several representative temperatures.  Absorptions in the transmission are dissipative and therefore correspond to peaks in k.  Dramatic changes in the spectra are observed across the zero field transitions T$_{\text{SR}}$ $\approx$ 37 K and T$_{\text{Ho}}$ $\approx$ 5.25K. }
\label{SIFig4}
\end{figure}

Here we display the full temperature dependence of the crystal field spectra of HMO.  \autoref{SIFig3} displays the magnitude of the complex transmission as a function of frequency and temperature for measurement geometries (a) $\vec{\text{h}}_{\text{ac}}$ $\parallel$ $c$ and (b) $\vec{\text{h}}_{\text{ac}}$ $\perp$ $c$ respectively.  One can show that the axial symmetry of the lattice constrains the transmission matrix such that only these two orientations give unique responses in zero field.  The complex transmissions were then numerically inverted to extract the complex index of refraction of the sample.  \autoref{SIFig4} displays the extracted imaginary part of the index of refraction for both zero field orientations at several representative temperatures.  These are the data that were used to generate the image plots shown in Figure 2 of the main text.  One can see that the spectra are dominated by strong dissipation due to Ho ($^5$I$_8$) crystal field excitations below 1 THz and around 1.5 THz.  The excitation at 1.3 THz in the $\vec{\text{h}}_{\text{ac}}$ $\perp$ $c$ orientation is the Mn AFR.

One can see from \autoref{SIFig4} that the crystal field spectra becomes considerably complex as the temperature is reduced, strongly departing from the simplistic picture of singlets and doublets outlined above.  Many crystal field excitations split into clusters of closely packed and highly overlapping excitations at low temperatures.  In order to better understand the crystal field spectrum, the imaginary part of the index of refraction were fit to a generic model of Lorentzian oscillators on a linear background.  Here our analysis will be restricted to only the crystal field excitation energies as their intensities and widths are difficult to extract with so many overlapping excitations.  

\begin{figure}
\includegraphics[width=0.8\columnwidth, keepaspectratio]{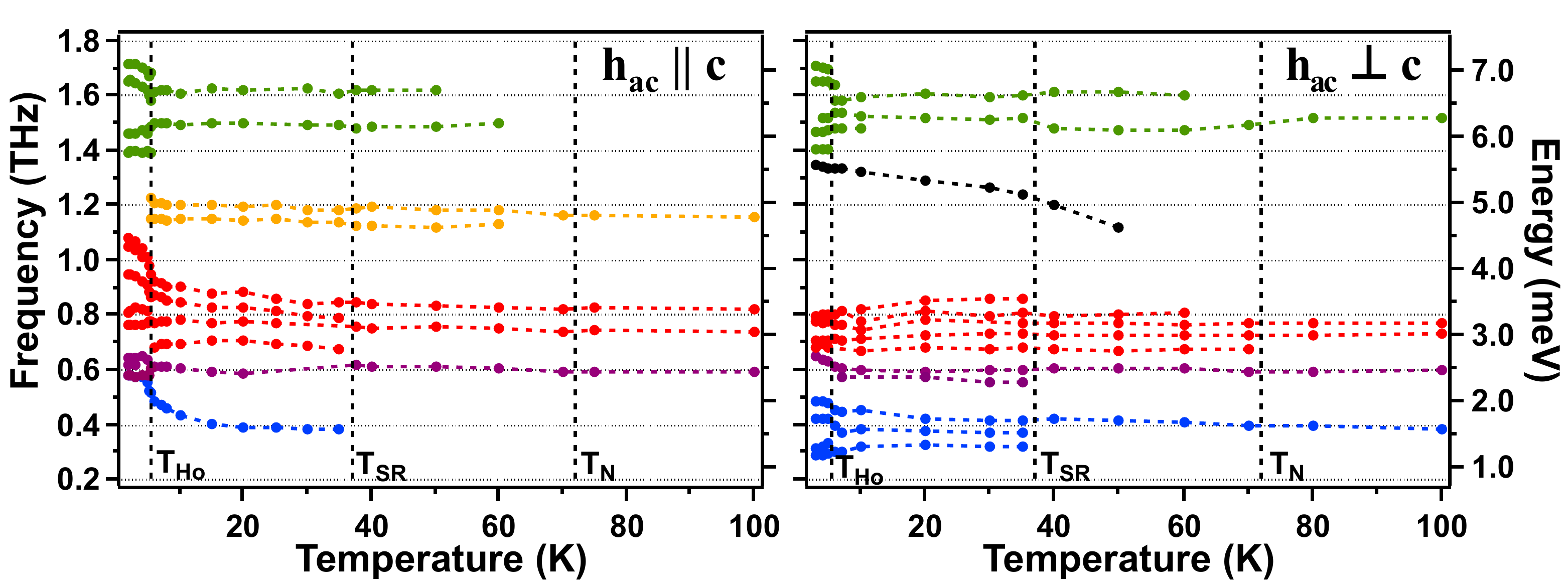}
\caption{Temperature dependence of the infrared excitation energies of HMO derived from fitting the imaginary part of the index of refraction for the (left) $\vec{\text{h}}_{\text{ac}}$ $\parallel$ c and (right) $\vec{\text{h}}_{\text{ac}}$ $\perp$ c orientations.  Color denotes each family of excitations.  The black curve in the $\vec{\text{h}}_{\text{ac}}$ $\perp$ c orientation is the Mn AFR.  Vertical dashed lines mark the three transition temperatures T$_{\text{N}}$, T$_{\text{SR}}$, and T$_{\text{Ho}}$.}
\label{SIFig5}
\end{figure}

Figure \ref{SIFig5} displays the excitation energies as a function of temperature for both measurement geometries.  Vertical dashed lines mark the three zero field transition temperatures T$_{\text{N}}$, T$_{\text{SR}}$, and T$_{\text{Ho}}$.  Color denotes distinct crystal field excitations many of which split into a multiplet of nearly degenerate excitations as the temperature is reduced. Although new excitations develop at both T$_{\text{SR}}$ and T$_{\text{Ho}}$, the energy spectrum shown in Figure \ref{SIFig5} shows relatively weak temperature dependence above T$_{\text{Ho}}$.  This is consistent with optical spectroscopy of other hexagonal manganites \cite{Standard2012} and the related compound HoMn$_2$O$_5$ \cite{Sirenko2008} which observe that the energy eigenvalues of the crystal field spectrum to be nearly independent of magnetic ordering.  Instead, magnetic exchange manifests itself was a redistribution of spectral weight between optically active excitations.  This has previously been explained as stemming from the change in symmetry of the ligand field at the rare-earth site as new magnetic order develops, which modifies the selection rules of the system, shifting spectral weight between excitations. 

However, unlike the transitions at T$_{\text{N}}$ and T$_{\text{SR}}$, the spectra at T$_{\text{Ho}}$ displays a dramatic renormalization of the crystal field excitation energies.  This low temperature energy renormalization, which has not been previously identified, presumably stems from magnetic exchange interactions in HMO.  Whether such exchange is between Ho-Ho or Ho-Mn moments remains an open question.  However, Ho-Mn interactions are expected to be roughly an order of magnitude stronger than Ho-Ho exchange and as we demonstrated in the main text, Ho-Mn interactions are particularly strong in HMO.  Therefore, we speculate that such a renormalization of the Ho crystal field spectrum below T$_\text{Ho}$ may stem from Ho-Mn exchange but further investigation is required to confirm.  Table \ref{CF_Table} summarizes our characterization of the low energy excitation spectrum of HMO.  Excitation energies, reported for temperatures T$>$ T$_\text{SR}$, are shown as well as a comparison to previously reported values, although several excitations are reported here for the first time.   Additonally shown are the selection rules by measurement orientation for each excitation.  

\begin{table*}[h]
\centering
\setlength{\tabcolsep}{3pt}
\begin{tabular}{|c|c|c|c|c|}
\hline\hline
Excitation & \begin{tabular}{@{}c@{}} Energy \\(T $>$ T$_{\text{SR}}$) \end{tabular} & \begin{tabular}{@{}c@{}}  Previously \\Reported Energies \end{tabular}   & \begin{tabular}{@{}c@{}}  Optically Active \\ h$_\text{ac}$ $\parallel$ c \end{tabular} & \begin{tabular}{@{}c@{}} Optically Active \\ h$_\text{ac}$ $\perp$ c \end{tabular}   \\
\hline

CF 1 & 1.51 meV & 1.48 meV (Ref. \cite{Vajk2005}) & \begin{tabular}{@{}c@{}} T$_\text{Ho}<$T$<$T$_\text{SR}$ \end{tabular} & Always Active \\ \hline

CF 2 & 2.43 meV & 2.41 meV (Ref. \cite{Vajk2005}) & \begin{tabular}{@{}c@{}} T$>$T$_\text{SR}$ \\ T$<$T$_\text{Ho}$ \end{tabular} & Always Active \\ \hline

CF 3 & 3.09 meV \& 3.40 meV & \begin{tabular}{@{}c@{}} 3.13 meV (Ref. \cite{Vajk2005}) \\3.40 meV (Ref. \cite{Talbayev2008}) \end{tabular} & Always Active & Always Active \\ \hline

CF 4 & 4.68 meV \& 4.89 meV & 4.84 meV (Ref. \cite{Talbayev2008}) & T$>$T$_\text{Ho}$ & Never Active\\ \hline

CF 5 &  6.17 meV \& 6.71 meV & 6.46 meV (Ref. \cite{Talbayev2008}) & Always Active & Always Active\\ \hline

\end{tabular}
\caption{Summary of our characterization of the crystal field spectra of HMO.  Crystal field excitation energies are reported at temperatures T $>$ T$_\text{SR}$.  Also included are previously reported crystal field energies found via neutron scattering \cite{Vajk2005} and infrared spectroscopy \cite{Talbayev2008}. Values from Ref. \onlinecite{Talbayev2008} were reported at T = 10K.  CF 4 was observed in Ref. \onlinecite{Talbayev2008} but it's energy was not reported.  The value listed in the table is estimated from figure 1 of Ref. \onlinecite{Talbayev2008}.  Several crystal field excitations are reported here for the first time.}
\label{CF_Table}
\end{table*}

\section{Full Magnetic Field Dependence}

In the main text, data was shown only in the orientation in which the THz oscillatory magnetic field $\vec{\text{h}}_{\text{ac}}$ $\perp$ c and the applied dc magnetic field $\vec{\text{H}}_{\text{dc}}$ $\parallel$ c and only at 20 K.  In this section, we display the full data set for this orientation as well as other orientations obtained by varying the direction of $\vec{\text{h}}_{\text{ac}}$ and $\vec{\text{H}}_{\text{dc}}$ with respect to the c axis.  
\pagebreak

\subsection{$\vec{\text{h}}_{\text{ac}}$ $\perp$ c, $\vec{\text{H}}_{\text{dc}}$ $\parallel$ c}

\begin{figure}[hp]
\includegraphics[width=0.95\columnwidth, keepaspectratio]{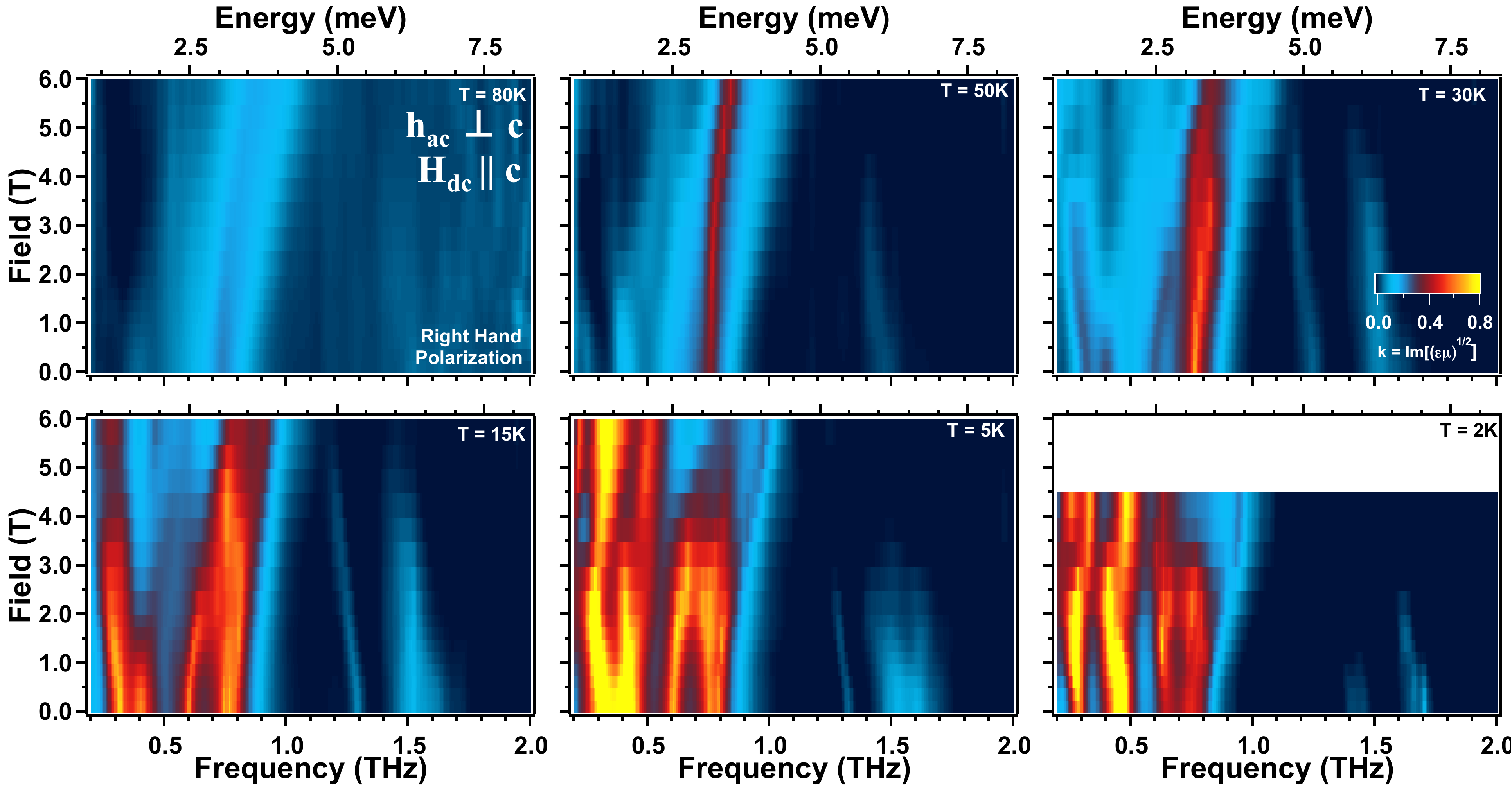}
\caption{Image plots of the imaginary part of the index of refraction with the THz oscillatory magnetic field $\vec{\text{h}}_{\text{ac}}$ $\perp$ c and the applied dc magnetic field $\vec{\text{H}}_{\text{dc}}$ $\parallel$ c at several representative temperatures.  Data is presented in the in the right hand channel of the circular basis.}
\label{SIFig6}
\end{figure}

\begin{figure*}[h!]
\includegraphics[width=0.95\columnwidth, keepaspectratio]{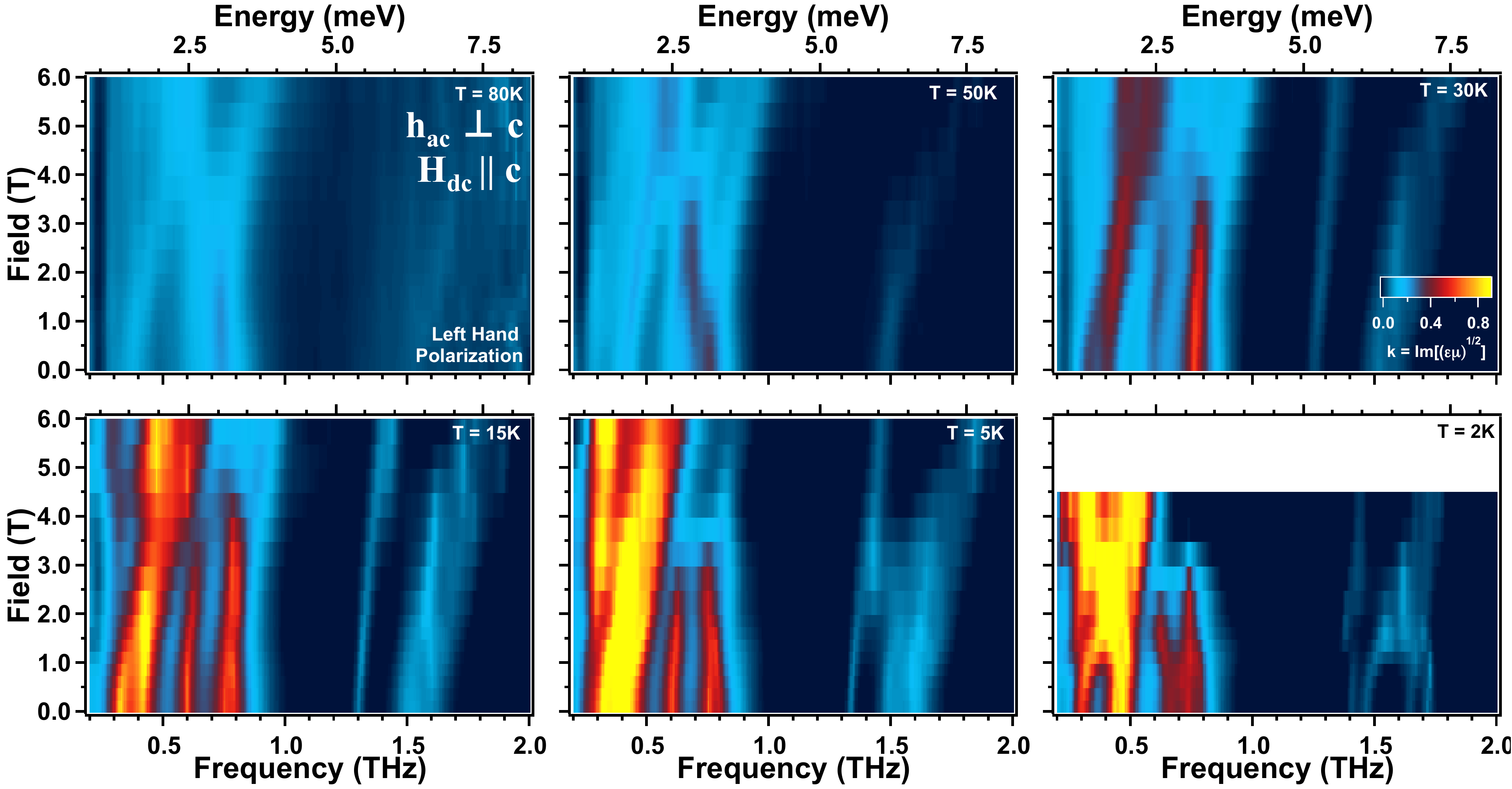}
\caption{Image plots of the imaginary part of the index of refraction with the THz oscillatory magnetic field $\vec{\text{h}}_{\text{ac}}$ $\perp$ c and the applied dc magnetic field $\vec{\text{H}}_{\text{dc}}$ $\parallel$ c at several representative temperatures.  Data is presented in the in the left hand channel of the circular basis.}
\label{SIFig7}
\end{figure*}
\pagebreak

\subsection{$\vec{\text{h}}_{\text{ac}}$ $\parallel$ c, $\vec{\text{H}}_{\text{dc}}$ $\perp$ c}

\begin{figure*}[h]
\includegraphics[width=0.90\columnwidth, keepaspectratio]{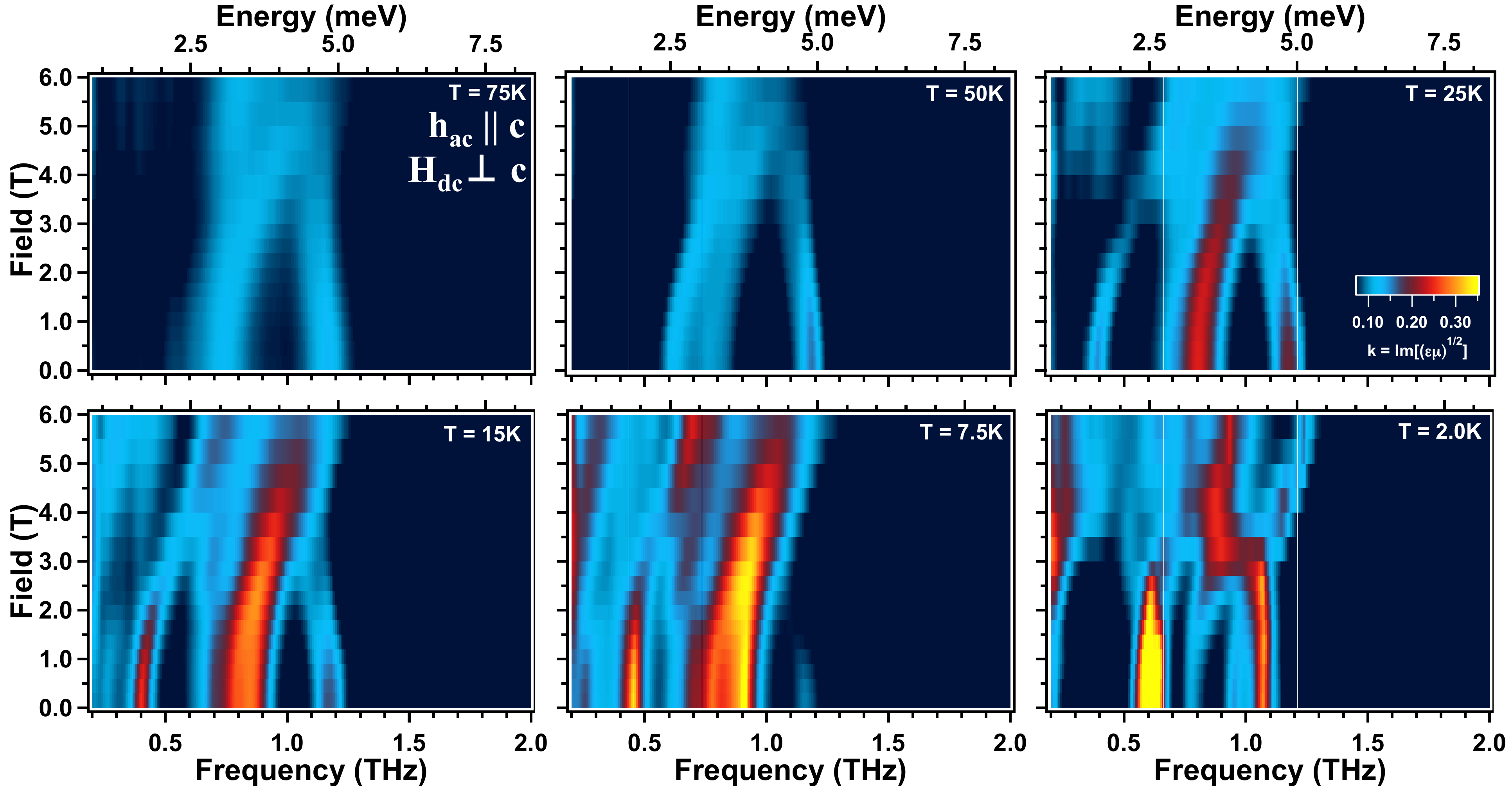}
\caption{Image plots of the imaginary part of the index of refraction with the THz oscillatory magnetic field $\vec{\text{h}}_{\text{ac}}$ $\parallel$ c and the applied dc magnetic field $\vec{\text{H}}_{\text{dc}}$ $\perp$ c at several representative temperatures.}
\label{SIFig8}
\end{figure*}

\subsection{$\vec{\text{h}}_{\text{ac}}$ $\perp$ c, $\vec{\text{H}}_{\text{dc}}$ $\perp$ c}

\begin{figure*}[h]
\includegraphics[width=0.90\columnwidth, keepaspectratio]{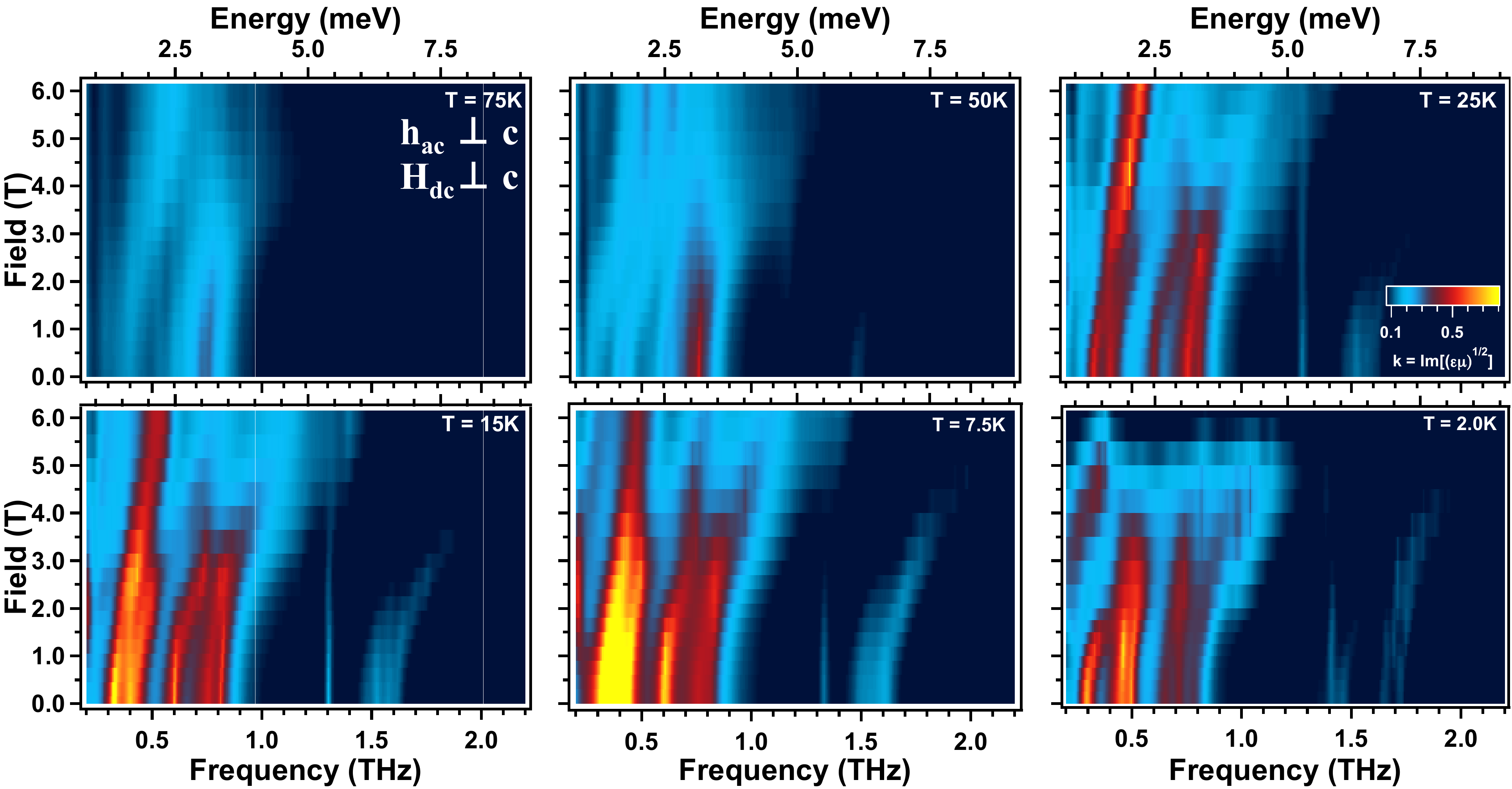}
\caption{Image plots of the imaginary part of the index of refraction with the THz oscillatory magnetic field $\vec{\text{h}}_{\text{ac}}$ $\perp$ c and the applied dc magnetic field $\vec{\text{H}}_{\text{dc}}$ $\perp$ c at several representative temperatures.}
\label{SIFig9}
\end{figure*}


\bibliography{HoMnO3_Mag}

\end{document}